\documentclass[12pt]{article}
\usepackage{bezier}
\usepackage{amssymb}
\usepackage{epsfig}

\newcommand{\nc}{\newcommand}
\nc{\mlgraph}{{multiple-line graph }}
\nc{\mlgraphs}{{mul\-tiple\--line graphs }}
\nc{\Mlgraph}{{Multiple-line graph }}

\nc{\mldiagram}{{multiple-line diagram }}
\nc{\mldiagrams}{{multiple-line diagrams }}

\nc{\mlmoment}{{multiple-line moment}}
\nc{\mlmoments}{{multiple-line moments}}

\nc{\be}{\begin{equation}}
\nc{\ee}{\end{equation}}
\nc{\bea}{\begin{eqnarray}}
\nc{\eea}{\end{eqnarray}}
\nc{\bela}{\begin{eqnarray*}}
\nc{\eela}{\end{eqnarray*}}

\nc{\eqn}[1]{{(\ref{#1})}}
\nc{\cA}{{\cal A}}
\nc{\cB}{{\cal B}}
\nc{\cC}{{\cal C}}
\nc{\cD}{{\cal D}}
\nc{\cE}{{\cal E}}
\nc{\cF}{{\cal F}}
\nc{\cG}{{\cal G}}
\nc{\cH}{{\cal H}}
\nc{\cI}{{\cal I}}
\nc{\cJ}{{\cal J}}
\nc{\cK}{{\cal K}}
\nc{\cL}{{\cal L}}
\nc{\cM}{{\cal M}}
\nc{\cN}{{\cal N}}
\nc{\cO}{{\cal O}}
\nc{\cP}{{\cal P}}
\nc{\cQ}{{\cal Q}}
\nc{\cR}{{\cal R}}
\nc{\cS}{{\cal S}}
\nc{\cT}{{\cal T}}
\nc{\cU}{{\cal U}}
\nc{\cV}{{\cal V}}
\nc{\cW}{{\cal W}}
\nc{\cX}{{\cal X}}
\nc{\cY}{{\cal Y}}
\nc{\cZ}{{\cal Z}}

\nc{\simo}[1]{{\stackrel{#1}{\simeq}}}
\nc{\geqo}[1]{{\stackrel{#1}{\geq}}}
\nc{\geo}[1]{{\stackrel{#1}{>}}}
\nc{\guo}[1]{{\stackrel{#1}{\succ}}}

\nc{\rbo}{\raisebox}
\nc{\RR} {\rangle \! \rangle}
\nc{\LL} {\langle \! \langle}
\nc{\rmi}[1]{{\mbox{\small #1}}}
\nc{\eq}{eq.~}
\nc{\nr}[1]{(\ref{#1})}
\nc{\ul}{\underline}
\nc{\mc}{\multicolumn}
\nc{\todo}[1]{\par\noindent{\bf $\rightarrow$ #1}}

\nc{\cu}{{\cal u}}

\title{
  \begin{flushright} {\small $\begin{array}{ l } \mbox{HD--THEP--98--11} \\
    \mbox{WUB--98--09} \end{array} $}
 \end{flushright}
Dynamical Linked Cluster Expansions: \\
A Novel Expansion Scheme for Point-Link-Point-Interactions}

\author{Hildegard~Meyer-Ortmanns\thanks{E-mail address:
ortmanns@theorie.physik.uni-wuppertal.de}
        \\
\small  Institut
\small  f\"ur Theoretische Physik\\
\small  Bergische Universit\"at Wuppertal \\
\small  Gau\ss strasse 20 \\
\small  D-42097 Wuppertal, Germany
  \\ 
\small and \\
 Thomas~Reisz\thanks{Supported by a Heisenberg Fellowship,
      E-mail address: reisz@thphys.uni-heidelberg.de}
        \\
\small  Institut
\small  f\"ur Theoretische Physik\\
\small  Universit\"at Heidelberg \\
\small  Philosophenweg 16 \\
\small  D-69120 Heidelberg, Germany}
       
\begin{document}

\maketitle

\begin{abstract}
\setlength{\baselineskip}{1pt}
Dynamical linked cluster expansions are linked cluster expansions with
hopping parameter terms endowed with their own dynamics. 
This amounts to a generalization from 2-point to point-link-point
interactions. We develop an
associated graph theory with a generalized notion of connectivity and
describe an algorithmic generation of the new \mlgraphs. 
We indicate physical
applications to spin glasses, partially annealed neural networks and SU(N)
gauge Higgs systems. In
particular the new expansion  technique provides the possibility of avoiding
the replica-trick in spin glasses. 
We consider variational estimates for the SU(2) Higgs
model of the electroweak phase transition. The results for the
transition line,
obtained by dynamical linked cluster expansions, agree
quite well with corresponding high precision Monte Carlo results.

\end{abstract}


%
%
%
%
\section{Introduction}

Linked cluster expansions (LCEs) have a long tradition in statistical
physics. Originally applied to classical fluids, later to magnetic systems
(\cite{wortis},\cite{itzykson},\cite{guttmann} and references therein),
they were generalized to applications in particle physics in the eighties
\cite{LW1}. There they have been used to study the
continuum limit of a lattice $\Phi^4$ field theory in 4 dimensions at zero 
temperature. In \cite{thomas1,thomas2}
they were further generalized to field
theories at finite temperature, simultaneously the highest 
order in the expansion parameter was increased to 18.
Usually the analytic expansions are obtained as graphical expansions.
Because of the
progress in computer facilities and the development of efficient
algorithms for generating the graphs, it is nowadays possible to handle
of the order of billions of graphs. The whole range
from high temperatures down to the critical region becomes available,
and thermodynamic quantities like critical indices and critical temperatures
are determined with high precision
(the precision is comparable or even better than in corresponding
high quality Monte Carlo results)
\cite{thomas2}-\cite{butera}.
An extension
of LCEs to a finite volume in combination with a high order in the 
expansion parameter turned out to be a particularly powerful tool for
investigating the phase structure of systems with first and second order
transitions by means of a finite size scaling analysis \cite{hilde1}.

Linked cluster expansions are series expansions of the free energy 
and connected correlation functions about an
ultralocal, decoupled theory in terms of a hopping parameter $K$.
The corresponding graphical representation is a sum in terms of connected
graphs. The value of $K$ parametrizes the strength of interactions
between fields at different lattice sites. Usually they are
chosen as nearest neighbours, but also more general, less local couplings
lead to convergent expansions (\cite{pordt1} and references
therein) under appropriate conditions on the decay of the interactions.
In contrast to the ultralocal
terms of a generic interaction we will sometimes refer to hopping 
terms as non-ultralocal.

In this paper we develop dynamical linked cluster expansions (DLCEs).
These are linked cluster expansions with  hopping parameter terms
that are endowed with their own dynamics. 
Such systems are realized
in spin glasses or partially annealed neural networks with (fast) spins
and (slow) interactions \cite{sherrington}-\cite{penney}. 
They also occur in variational
estimates for SU(N)-gauge-Higgs systems as we will show 
later in this paper. Like LCEs
they are expected to converge for a large class of interactions.

Formally DLCEs amount to a generalization of an expansion scheme 
from 2-point to point-link-point-interactions. 
These are interactions between
fields associated with two points and with one pair of
points called link. The points and links are not
necessarily embedded on a lattice, and the links need not be
restricted to nearest neighbours.
We develop a new \mlgraph theory in which a generalized notion of
connectivity plays a central role.
Standard notions of equivalence classes of graphs like
1-line irreducible and 1-vertex irreducible graphs have to be generalized,
and new notions like 1-multiple-line irreducible graphs must be
defined in order to give a systematic classification.
We describe algorithms for generating these classes. 
This is essential, since the number of graphs rapidly increases 
with the order in the expansion parameter.
It typically exceeds the number of LCE-graphs by more than 
an order of magnitude.

The paper is organized as follows. In section 2 we specify the
models that admit a DLCE. We introduce \mlgraphs 
and explain the idea behind the abstract notions of \mlgraph theory.
Detailed definitions of \mlgraphs, related notions and the
computation of weights are given in section 3.
Section 4 treats the issue of renormalization in the sense of
suitable resummations of graphs.
Algorithms for their generation are described in section 5.
Applications to spin glasses and
(partially annealed) neural networks are indicated in section 6. There it
is of particular interest that DLCEs allow for the possibility of 
avoiding the replica trick. In section 7 we present results for
the transition line of an SU(2) Higgs model.
These are obtained by DLCEs applied to gap equations that follow from
convexity estimates of the free energy density. 
The results are in good agreement with corresponding high precision 
Monte Carlo results \cite{jansen}.
Section 8 contains a summary and outlook. 

The reader who is more 
interested in our actual applications of DLCEs may skip sections 3-5 in a
first reading. These sections are of interest from a systematic point of
view and essential if he is interested in applying DLCEs.
%
%

%
%
\section{\label{models}A Short Primer to DLCEs}

In  this section we first specify the class of models for which we develop
dynamical linked cluster expansions. Next we illustrate some basic notions
of multiple-line graph theory, in particular the need for a new notion of
connectivity.

By $\Lambda_0$ we denote a finite or infinite set of points. One of its
realizations is a
hypercubic lattice in $D$ dimensions, 
infinite or finite in some directions with the topology of a torus.
$\Lambda_1$ denotes the set of unordered pairs $(x,y)$ of sites
$x,y\in\Lambda_0$, $x\not= y$, also called
unoriented links, and $\overline\Lambda_1$ a subset of $\Lambda_1$.

We consider physical systems with a partition function of the generic form
\bea\label{2.zgen}
  && Z(H,I,v) \; \equiv \; \exp{W(H,I,v)} \nonumber \\
  && \; = \;  \cN  \int \cD \phi \cD U
\exp{(-S(\phi,U,v))} \exp{(\sum_{x \in \Lambda_\circ}H(x)
          \phi(x) +\sum_{l \in \overline\Lambda_1}I(l)U(l))},
\eea
with measures
\be\label{2m}
    \cD\phi =  \prod_{x\in \Lambda_0}  d\phi(x)
\quad , \quad
 \cD U = \prod_{l \in \overline\Lambda_1} dU(l)
\ee
and action
\be\label{2act}
S(\phi,U,v) \; = \; \sum_{x \in \Lambda_\circ}S^\circ(\phi(x))
             +\sum_{l \in \overline\Lambda_1}S^1(U(l))
             - \frac{1}{2} \, \sum_{x,y \in \Lambda_\circ} 
               v(x,y)\phi(x)U(x,y)\phi(y) ,
\ee
with non-ultralocal couplings
\bea
 &&  v(x,y) \; = \; v(y,x) \; \not= \; 0 
 \qquad   \mbox{only for $(x,y)\in\overline\Lambda_1$}, \nonumber \\
 && \mbox{in particular} \; v(x,x) \; = \; 0.
\eea
For later convenience the normalization via $\cN$ is chosen such that
$W[0,0,0]=0$.

The field $\phi(x)$ is associated with the sites $x\in\Lambda_0$
and the field $U(l)$
lives on the links $l\in\overline\Lambda_1$,
and we write $U(x,y)=U(l)$ for $l=(x,y)$.
For definiteness and for simplicity of the notation here we assume
$\phi(x) \in {\bf R}$ and $U(l) \in {\bf R}$.
In our actual applications $\phi$ is a
4-component scalar field and $U$ an SU(2)-valued gauge field,
or $\phi$ and $U$ both are Ising spins, or the $\phi$s are the (fast)
Ising spins and the $U$s $\in {\bf R}$ the (slow) interactions.
The action is split into two ultralocal parts, $S^\circ$ depending
on fields on single sites, and $S^1$ depending on fields
on single links $l \in \overline\Lambda_1$.
For simplicity we choose $S^1$ as the same function for all
links $l\in\overline\Lambda_1$. 
We may identify $\overline\Lambda_1$ with the support of $v$,
\be
  \overline\Lambda_1 \; = \; \{ l=(x,y) \; \vert \;
  v(x,y) \not=0 \}.
\ee
The support of $v(x,y)$ need not be restricted to nearest neighbours,
also the precise form of $S^\circ$ and $S^1$
does not matter
for the generic description of DLCEs, $S^\circ$ and $S^1$ can be any
polynomials in $\phi$ and $U$, respectively.
The only restriction is 
the existence of the partition function.
In one of our
applications $S^\circ$ will be a $\phi^4$-type theory
with O(N) symmetry,
$S^1$ consists of a term linear or quadratic in $U$.

Note that the
interaction term $v(x,y)~\phi(x)~U(x,y)~\phi(y)$ contains a
point-link-point-interaction and generalizes the 2-point-interactions 
$v(x,y)~\phi(x)~\phi(y)$ of usual hopping parameter expansions.
The effective coupling of the $\phi$ fields has its own dynamics
governed by $S^1(U)$, the reason why we have called our new
expansion scheme {\it dynamical} LCE.

Dynamical linked cluster expansions are induced from a Taylor expansion of
$W(H,I,v)=\ln Z(H,I,v)$ about $v=0$, the limit of a completely decoupled
system.
We want to express the series for $W$ in terms of connected graphs.
Let us consider the generating equation
\bea \label{2.genequation}
\partial W/\partial v(xy) & = & 1/2 <\phi(x) U(x,y) \phi(y)> \nonumber \\
      & = & 1/2 
   \biggl( W_{H(x) I(x,y) H(y)} + W_{H(x)H(y)}W_{I(x,y)} \nonumber \\
            & + & W_{H(x)I(x,y)}W_{H(y)}  \nonumber
         +W_{I(x,y)H(y)}W_{H(x)}  \nonumber   \\
      & + & W_{H(x)}W_{H(y)}W_{I(x,y)} 
   \biggr) .
\eea

Here $<\cdot>$ denotes the normalized expectation value
w.r.t. the partition function of Eq.~(\ref{2.zgen}). Subscripts $H(x)$ and
$I(x,y)=I(y,x)=I(l)$ denote the derivatives of $W$ w.r.t.
$H(x)$ and $I(x,y)$, respectively.

Next we would like to represent the right hand side
of Eq.~(\ref{2.genequation}) 
in terms of connected graphs.
Once we have such a representation for the first derivative of $W$ w.r.t. 
$v$, grapical expansions for the higher derivatives can be traced back
to the first one. 

For each $W$ in Eq.~(\ref{2.genequation}) we draw a shaded bubble,
for each derivative w.r.t. $H$ a solid line, called a $\phi$-line, 
with endpoint vertex $x$,
and for each derivative w.r.t. $I$ a dashed line,
called a $U$-line, with link label $l=(x,y)$. 
The main graphical constituents are
shown in Fig.~1. Two $\phi$-lines with endpoints $x$ and $y$ are
then joined by means of a dashed $U$-line with label $l$, 
if the link $l$ has $x$
and $y$ as its endpoints, i.e. $l=(x,y)$. According to these rules
Eq.~(\ref{2.genequation}),
multiplied by $v(x,y)$ and summed over $x$ and $y$,
is represented by Fig.~2.
Note that, because of the Taylor operation, each solid line from $x$
to $y$ carries a factor $v(x,y)$.

Since the actual need for a new type of connectivity is not 
quite obvious
from Fig.~2, because Eq.~(\ref{2.genequation}) does not contain higher
than first order derivatives w.r.t. $I$, let us consider a term

\begin{figure}[ht]

\begin{center}
\setlength{\unitlength}{0.8cm}

%
%
\begin{picture}(15.0,3.0)

%
%
%

\epsfig{bbllx=-333,bblly=324,
        bburx=947,bbury=684,
        file=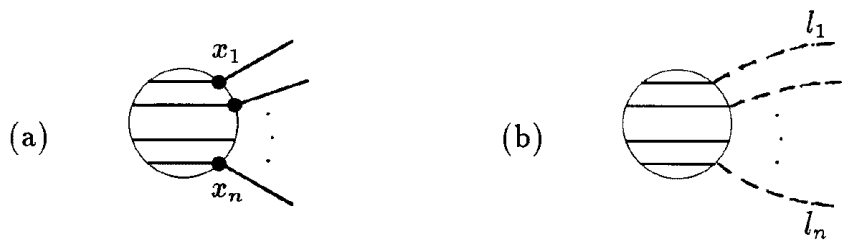,
        scale=0.27}

%
%

\end{picture}
%
%
\end{center}

\caption{\label{bubbles} Graphical representation of the derivatives
of $W(H,I,v)$.
(a) $n$-point function 
${\partial^n W}/{\partial H(x_1) \cdots \partial H(x_n)}$,
(b) $n$-link function
${\partial^n W}/{\partial I(l_1) \cdots \partial I(l_n)}$.
}
\end{figure}

\begin{figure}[ht]

\begin{center}
\setlength{\unitlength}{0.8cm}

%
%
\begin{picture}(15.0,7.0)

%
%
%

\epsfig{bbllx=-333,bblly=139,
        bburx=947,bbury=797,
        file=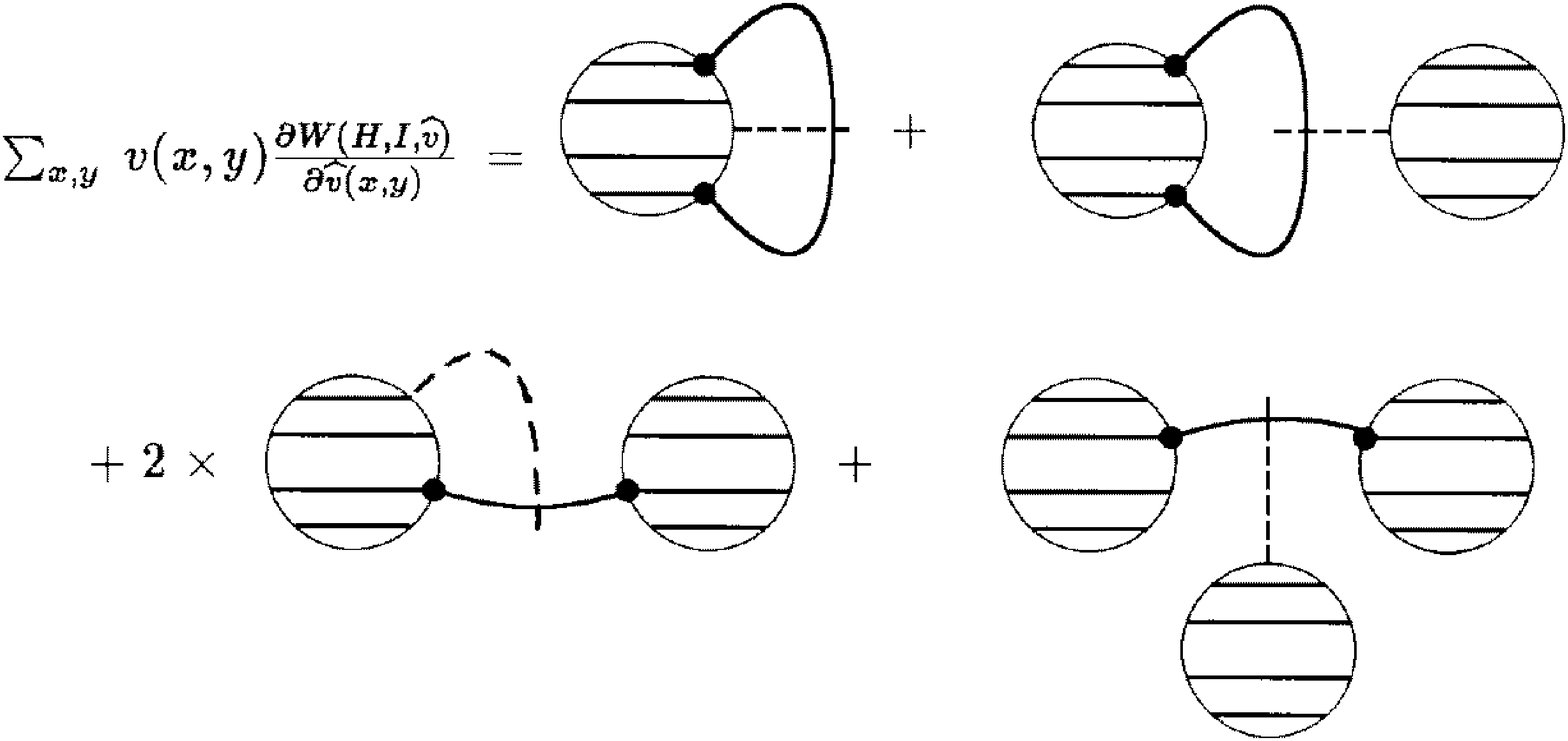,
        scale=0.27}

%
%

\end{picture}
%
%
\end{center}

\caption{\label{generation} Generating equation of the graphical
expansion of DLCEs.
The solid line in each graph carries a propagator $v(x,y)$.
A dashed $U$-line with label $l$ intersects a solid line with
endpoints $x$ and $y$ if $l=(x,y)$.}
\end{figure}

\be
W_{H(x)}W_{H(y)}W_{H(r)}W_{H(s)}
W_{I(x,y)I(r,s)}
\ee
occurring in the second derivative of $W$ w.r.t. $v(x,y)$, $v(r,s)$.
According to the above rules this term would be represented as shown in
Fig. 3a. While the 2 vertices in the last term of
Fig. 2 are connected
in the usual sense via a common (solid) line (the dashed line
with an attached bubble could be omitted in this case), the graph in 
Fig. 3a would be disconnected in the old sense,
since neither $x$ nor $y$ are line-connected with $r$ and $s$, but -as a
new feature of DLCE graphs- the lines from $x$ to $y$ and from $r$ to $s$
are connected via the dashed lines emerging from a common bubble shown
in the middle of the graph.
As we see from Fig. 3a, we need an additional notion of
connectivity referring to the possibility of multiple-line connectivity.
While the analytic expression is fixed, 
it is a matter of convenience
to further simplify the graphical notation of Fig.~3a 
at $v=0$. Two possibilities are shown in Fig.~3b
and Fig.~3c. To Fig. 3b we later refer in the formal definition
of the new type of multiple-line connectivity.
In the familiar standard notion of connectivity two vertices of a graph are
connected via lines. The vertices are line-connected.
Already there, in a dual language, one could call two
lines connected via vertices. The second formulation is just appropriate
for our need to define when two lines are connected. The corresponding
vertices mediating the connectivity of lines are visualized
by tubes, in Fig. 3b we have just one of them.
The tubes should be distinguished from the former type of
vertices represented as full dots which are connected via bare 
$\phi$-lines.
In Fig.~3c we show a simplified representation of Fig.~3b
that we actually use in graphical expansions.

\begin{figure}[h]

\begin{center}
\setlength{\unitlength}{0.8cm}

%
%
\begin{picture}(15.0,7.0)

%
%
%

\epsfig{bbllx=-333,bblly=186,
        bburx=947,bbury=606,
        file=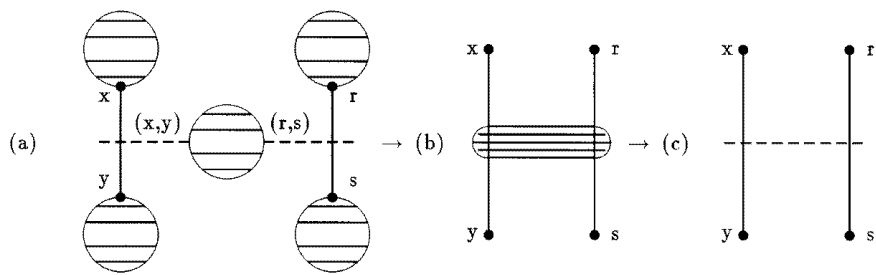,
        scale=0.27}

%
%

\end{picture}
%
%
\end{center}

\caption{\label{mlines} Representation of $W_{H(x)} W_{H(y)}
W_{H(r)} W_{H(s)} W_{I(x,y) I(r,s)}$.
(a) according to the rules of Fig.~1 and 2,
(b) same as (a), but at $v=0$ and simplified for a formal definition
of multiple-line connectivity, cf.~section 3,
(c) same as (b), but for use in the actual graphical
representations.}
\end{figure}

The derivative terms have to be evaluated at $v=0$. For $v=0$ we have
a decomposition of $W$ according to
\be
W(H,I,v=0)= \sum_{x \in \Lambda_0} W^\circ(H(x))
    +\sum_{l \in \overline\Lambda_1} W^1(I(l))
\ee
with 
\be\label{2.w0}
\exp{W^\circ(H)} \; \equiv \; Z^\circ(H) =
 \frac{\int^\infty_{-\infty} d\phi
 \exp{(-S^\circ(\phi)+H\phi)}}{\int^\infty_{-\infty} d\phi
 \exp{(-S^\circ(\phi))}}
\ee
and
\be\label{2.w1}
\exp{W^1(l)} \; \equiv \; Z^1(I) = \frac{\int^\infty_{-\infty} dU
\exp{(-S^1(U)+IU)}}{\int^\infty_{-\infty} dU
\exp{(-S^1(U))}} .
\ee
In Eq.s~(\ref{2.w0},\ref{2.w1}) we have omitted any single site or single
link dependence, because we assume that $S^\circ$ and 
$S^1$ are the same for all $x \in \Lambda_0$ and
$l\in \overline{\Lambda_1}$, respectively.
Therefore, at $v=0$, the only non-vanishing derivatives of $W$ are 
\be\label{2.wd1}
   \left. W_{H(x_1)H(x_2)...H(x_n)} \right\vert_{v=0} \; = \;
   \frac{\partial^n W^\circ(H(x_1))}
        {\partial H(x_1)^n} \cdot \delta_{x_1,x_2,...x_n}
\ee
and
\be\label{2.wd2}
   \left. W_{I(l_1)I(l_2)...I(l_m)} \right\vert_{v=0} \; = \;
   \frac{\partial^m W^1(I(l_1))}
        {\partial I(l_1)^m} \cdot \delta_{l_1,l_2,...l_m},
\ee
but mixed derivatives w.r.t.~H and I vanish.
As anticipated in Fig.s~3b and 3c, for $v=0$ we replace the dashed bubbles
and graphically distinguish between bubbles with $\phi$-lines
and bubbles with $U$-lines. We define
\be
  \left. 
{
\setlength{\unitlength}{0.8cm}
%
%
\begin{picture}(3.0,1.0)

%

\put(0.0,0.0){
\setlength{\unitlength}{0.8cm}
\begin{picture}(4.0,0.0)

\put(0.0,0.0){\circle*{0.16}}
\qbezier(0.0,0.0)(1.0,0.8)(2.0,0.8)
\qbezier(0.0,0.0)(1.0,0.5)(2.0,0.5)
\qbezier(0.0,0.0)(1.0,-0.7)(2.0,-0.7)

\put(1.2,0.2){\makebox(0.2,0){$\cdot$}}
\put(1.2,-0.1){\makebox(0.2,0){$\cdot$}}
\put(1.2,-0.4){\makebox(0.2,0){$\cdot$}}

\end{picture}
}
%
    
\end{picture}
}
    \right\} \; n
    \qquad = \; v^{\circ c}_n \; = \;
    \left( \frac{\partial^n W^\circ(H)}{\partial H^n} \right)_{H=0} 
\ee
for a connected n-point vertex with $n \ge 1$ bare $\phi$-lines
emerging from it and 
\be
  \left. 
{
\setlength{\unitlength}{0.8cm}
%
%
\begin{picture}(4.0,1.0)

%

\put(0.0,-0.6){
\setlength{\unitlength}{0.8cm}
\begin{picture}(4.0,0.0)

\qbezier(1.0,1.5)(2.0,2.1)(3.0,1.5)
\qbezier(1.0,1.0)(2.0,1.6)(3.0,1.0)
\qbezier(1.0,0.0)(2.0,0.6)(3.0,0.0)

\put(1.6,1.1){\makebox(0.2,0){$\cdot$}}
\put(1.6,0.8){\makebox(0.2,0){$\cdot$}}
\put(1.6,0.5){\makebox(0.2,0){$\cdot$}}

\put(2.0,1.85){\line(0,1){0.2}}
\put(2.0,1.6){\line(0,1){0.2}}
\put(2.0,1.35){\line(0,1){0.2}}
\put(2.0,1.1){\line(0,1){0.2}}
\put(2.0,0.85){\line(0,1){0.2}}
\put(2.0,0.6){\line(0,1){0.2}}
\put(2.0,0.35){\line(0,1){0.2}}
\put(2.0,0.1){\line(0,1){0.2}}

\end{picture}
}
%

\end{picture}
}
   \right\} \; \nu
   \qquad = \; m^{1 c}_\nu \; = \;
   \left( \frac{\partial^\nu W^1(I)}{\partial I^\nu} \right)_{I=0}
\ee
for a connected $\nu$-line consisting of $\nu$ bare lines.
If $\nu=1$, we often omit the dashed line. If the bare lines
of a $\nu$-line are internal $\phi$-lines,
they get vertices attached to their endpoints, if they are external $U$-lines,
no vertices will be attached.
 
\vskip10pt

Let $V$ denote the lattice volume in $D$ dimensions.
The Taylor expansion of $W$ about $v=0$ to 
second order in $v$ then reads
\bea\label{2.long}
W(H,I,v) & = & W(H,I,v=0) \nonumber \\
 &+&\sum_{x,y \in \Lambda_0} v(x,y) \, \frac{1}{2} \, W_{H(x)}
W_{H(y)}W_{I(x,y)}  \nonumber \\
 & + & \frac{1}{2} \, \sum_{x,y,r,s \in \Lambda_0} 
  \frac{1}{4} \; v(x,y) v(r,s) 
 \nonumber \\ 
 \cdot 
 & \biggl( & 
    4 \, W_{H(y)}W_{H(s)}W_{H(r)H(x)}W_{I(x,y)}W_{I(r,s)} \nonumber \\
  &+& 2 \, W_{H(x)H(r)}W_{H(y)H(s)}W_{I(x,y)}W_{I(r,s)}  \nonumber \\
  &+& 4 \, W_{H(y)}W_{H(s)}W_{H(r)H(x)}W_{I(x,y)I(r,s)}  \nonumber \\
  &+& 2 \, W_{H(r)H(x)}W_{H(y)H(s)}W_{I(r,y)I(x,s)} \nonumber \\
  &+& W_{H(x)}W_{H(y)}W_{H(r)}W_{H(s)}W_{I(x,y)I(r,s)}
  \biggr)_{v=0} \nonumber \\
 &+& O(v^3),
\eea
where we have used that $v(x,x)=0$. For each $W$ in the products of
$W$s we now insert Eq.s~(\ref{2.wd1}),(\ref{2.wd2}).

If we choose $v$ in a standard way as next-neighbour
couplings 
\be
  v(x,y) = 2K \sum_{\mu=0}^{D-1} (\delta_{x+\hat{\mu},y}+
\delta_{x-\hat{\mu},y})
\ee
with $\hat{\mu}$ denoting the unit vector in $\mu$-direction,
Eq.~(\ref{2.long}) becomes in a graphical representation at $H=I=0$
\bea\label{2.graph}
  \frac{W(0,0,v)}{V} & = & (2K) \;\;
   \frac{1}{2} \; (2D) \;
%
{
\setlength{\unitlength}{0.8cm}
\begin{picture}(4.0,1.0)

\put(0.0,0.0){
\setlength{\unitlength}{1.0cm}
\begin{picture}(4.0,1.0)

\put(0.0,0.2){\circle*{0.16}}
\put(2.0,0.2){\circle*{0.16}}
\put(0.0,0.2){\line(1,0){2.0}}

\end{picture}
}

\end{picture}
%
}
 \nonumber \\
 &+& (2K)^2 \;\; \biggl\{ \frac{1}{2} \; (2D)^2 \;
%
{
\setlength{\unitlength}{0.8cm}
\begin{picture}(3.5,1.0)

\put(0.0,0.0){
\setlength{\unitlength}{1.0cm}
\begin{picture}(4.0,1.0)

\put(0.0,-0.2){\circle*{0.16}}
\put(0.0,0.6){\circle*{0.16}}
\put(2.0,0.2){\circle*{0.16}}
\qbezier(0.0,-0.2)(1.0,-0.2)(2.0,0.2)
\qbezier(0.0,0.6)(1.0,0.6)(2.0,0.2)

\end{picture}
}

\end{picture}
%
}
 \; + \; \frac{1}{4} \; (2D)
%
{
\setlength{\unitlength}{0.8cm}
\begin{picture}(4.0,1.0)

\put(0.0,0.0){
\setlength{\unitlength}{1.0cm}
\begin{picture}(4.0,1.0)

\put(0.0,0.2){\circle*{0.16}}
\put(2.0,0.2){\circle*{0.16}}
\qbezier(0.0,0.2)(1.0,-0.4)(2.0,0.2)
\qbezier(0.0,0.2)(1.0,0.8)(2.0,0.2)

\end{picture}
}

\end{picture}
%
}
 \nonumber \\
 &\;& \qquad\quad + \frac{1}{2} \; (2D)^2 \;
%
{
\setlength{\unitlength}{0.8cm}
\begin{picture}(3.5,1.0)

\put(0.0,0.0){
\setlength{\unitlength}{1.0cm}
\begin{picture}(4.0,1.0)

\put(0.0,-0.2){\circle*{0.16}}
\put(0.0,0.6){\circle*{0.16}}
\put(2.0,0.2){\circle*{0.16}}
\qbezier(0.0,-0.2)(1.0,-0.2)(2.0,0.2)
\qbezier(0.0,0.6)(1.0,0.6)(2.0,0.2)

\put(0.9,0.2){\line(0,1){0.15}}
\put(0.9,0.4){\line(0,1){0.15}}
\put(0.9,0.6){\line(0,1){0.15}}
\put(0.9,0.0){\line(0,1){0.15}}
\put(0.9,-0.2){\line(0,1){0.15}}
\put(0.9,-0.4){\line(0,1){0.15}}

\end{picture}
}

\end{picture}
%
}
 \; + \; \frac{1}{4} \; (2D)
%
{
\setlength{\unitlength}{0.8cm}
\begin{picture}(4.0,1.0)

\put(0.0,0.0){
\setlength{\unitlength}{1.0cm}
\begin{picture}(4.0,1.0)

\put(0.0,0.2){\circle*{0.16}}
\put(2.0,0.2){\circle*{0.16}}
\qbezier(0.0,0.2)(1.0,-0.4)(2.0,0.2)
\qbezier(0.0,0.2)(1.0,0.8)(2.0,0.2)

\put(1.0,0.2){\line(0,1){0.15}}
\put(1.0,0.4){\line(0,1){0.15}}
\put(1.0,0.6){\line(0,1){0.15}}
\put(1.0,0.0){\line(0,1){0.15}}
\put(1.0,-0.2){\line(0,1){0.15}}
\put(1.0,-0.4){\line(0,1){0.15}}

\end{picture}
}

\end{picture}
%
}
 \\
 &\;& \qquad\quad + \frac{1}{8} \; 2(2D) \;
%
{
\setlength{\unitlength}{0.8cm}
\begin{picture}(3.5,1.0)

\put(0.0,0.0){
\setlength{\unitlength}{1.0cm}
\begin{picture}(4.0,1.0)

\put(0.0,0.0){\circle*{0.16}}
\put(0.0,0.3){\circle*{0.16}}
\put(2.0,0.0){\circle*{0.16}}
\put(2.0,0.3){\circle*{0.16}}
\qbezier(0.0,0.0)(1.0,-0.5)(2.0,0.0)
\qbezier(0.0,0.3)(1.0,0.8)(2.0,0.3)

\put(1.1,0.2){\line(0,1){0.15}}
\put(1.1,0.4){\line(0,1){0.15}}
\put(1.1,0.6){\line(0,1){0.15}}
\put(1.1,0.0){\line(0,1){0.15}}
\put(1.1,-0.2){\line(0,1){0.15}}
\put(1.1,-0.4){\line(0,1){0.15}}

\end{picture}
}

\end{picture}
%
}
 \biggr\}
 \nonumber \\
 & + & O(K^3) . \nonumber
\eea

For clarity, here we have written explicitly the topological 
symmetry factors and the lattice embedding numbers. 
(Usually graphs represent their full weights including these factors.)
Note that the first two graphs of the second order contribution
also occur in a usual LCE with frozen
$U$-dynamics, the next two differ by an additional dashed 2-line 
and the last one becomes even
disconnected without the dashed line.

As usual, graphical expansions for correlation functions, in particular
susceptibilities, are generated from $W(H,I,v)$ by taking derivatives
w.r.t. the external fields $H$ and $I$.
Graphically this amounts to attaching external
$\phi$-lines and $U$-lines with
\be
{
\setlength{\unitlength}{0.8cm}
%
%
\begin{picture}(16.0,3.0)


\put(0.0,2.0){
\setlength{\unitlength}{0.8cm}
\begin{picture}(4.0,0.0)

\qbezier(0.0,0.0)(1.0,1.0)(2.0,1.0)
\put(2.0,1.0){\circle*{0.16}}

\put(3.2,0.5){\makebox(8.0,0){(1 endpoint) attached to vertices, e.g.}}
%

\qbezier(13.0,0.5)(14.0,0.5)(15.0,0.5)
\qbezier(13.0,0.5)(14.0,1.2)(15.0,0.5)
\qbezier(13.0,0.5)(14.0,-0.2)(15.0,0.5)
\put(13.0,0.5){\circle*{0.16}}
\put(15.0,0.5){\circle*{0.16}}
\qbezier(15.0,0.5)(15.5,0.5)(16.0,0.5)
\qbezier(13.0,0.5)(12.5,0.5)(12.0,0.0)

\put(14.0,-0.1){\line(0,1){0.15}}
\put(14.0,0.1){\line(0,1){0.15}}
\put(14.0,0.3){\line(0,1){0.15}}
\put(14.0,0.5){\line(0,1){0.15}}

\end{picture}
}
%


\put(0.0,0.0){
\setlength{\unitlength}{0.8cm}
\begin{picture}(16.0,0.0)

\qbezier(0.0,0.0)(1.0,1.0)(2.0,0.0)

\put(3.2,0.5){\makebox(8.0,0){(no endpoint) attached to $\nu$-lines, e.g.}}
%

\qbezier(13.0,0.5)(14.0,0.5)(15.0,0.5)
\qbezier(13.0,0.5)(14.0,1.2)(15.0,0.5)
\qbezier(13.0,0.5)(14.0,-0.2)(15.0,0.5)
\put(13.0,0.5){\circle*{0.16}}
\put(15.0,0.5){\circle*{0.16}}

\qbezier(13.0,-0.3)(14.0,0.2)(15.0,-0.3)
\put(14.0,-0.3){\line(0,1){0.15}}
\put(14.0,-0.1){\line(0,1){0.15}}
\put(14.0,0.1){\line(0,1){0.15}}
\put(14.0,0.3){\line(0,1){0.15}}
\put(14.0,0.5){\line(0,1){0.15}}

\end{picture}
}
%

\end{picture}
}
\ee
In passing we remark that the conventional LCE is included as a special case
of the DLCE, if the $U$-dynamics is "frozen" to some value $U_0\neq 0$,
so that
\bea
   W^1(I) & = &-S_1(U_0)+ I U_0, \nonumber \\
   \frac{\partial W^1(I)}{\partial I} & = & U_0, \\
   \frac{\partial^n W^1(I)}{\partial I^n} & = & 0
    \quad\mbox{for all $n>1$},  \nonumber
\eea
i.e., no n-lines do occur with $n>1$. In this case it becomes redundant to
attach dashed lines to bare lines. 
As mentioned above, in an LCE only the first three
contributions would be left in Eq.~(\ref{2.graph}).
%
%
%
%
\section{\label{basics}Graphical expansion}

%
%

\subsection{\Mlgraph theory}

The definition of a \mlgraph as it will be given here is adapted
to the computation of susceptibilities,
where the sum is taken over all possible locations of the fields.
The definition easily generalizes to correlation functions.

For details of the standard definiton of graphs in the framework of
linked cluster expansions and related notions
we refer  e.g.~to \cite{LW1,thomas1}.
Here, for convenience, we briefly recall the very definition of a graph
to point out the new properties
of \mlgraphs as defined below in this section.

A (standard LCE) graph or diagram is a structure
\be
  \widetilde\Gamma = (\widetilde\cL_\Gamma,
   \widetilde\cB_\Gamma,\widetilde E_\Gamma,\widetilde\Phi_\Gamma),
\ee
where $\widetilde\cL_\Gamma$ and $\widetilde\cB_\Gamma\not=\emptyset$
are disjoint sets
of internal lines and vertices of $\widetilde\Gamma$, respectively.
$\widetilde E_\Gamma$ is a map
\bea
  \widetilde E_\Gamma: \widetilde\cB_\Gamma & \to & \{0,1,2,\ldots\},
  \nonumber \\
   v & \to & \widetilde E_\Gamma(v)
\eea
that assigns to every vertex $v$ the number of external lines 
$\widetilde E_\Gamma(v)$ attached to it.
Finally, $\widetilde\Phi_\Gamma$ is the incidence relation
that assigns internal lines
to their two endpoints.

A \mlgraph or \mldiagram is a structure
\be
  \Gamma = (\cL_\Gamma, \cM_\Gamma, \cB_\Gamma,
             E_\Gamma^{(\phi)}, E_\Gamma^{(U)},
             \Phi_\Gamma, \Psi_\Gamma).
\ee
$\cL_\Gamma$, $\cM_\Gamma$ and $\cB_\Gamma$
are three mutually disjoint sets,
\bea
    \cL_\Gamma  &=& \mbox{set of bare internal lines of $\Gamma$}, \\
    \cM_\Gamma  &=& \mbox{set of multiple lines of $\Gamma$}, \\
    \cB_\Gamma  &=& \mbox{set of vertices of $\Gamma$}.
\eea
$E_\Gamma^{(\phi)}$ is a map
\bea
  E_\Gamma^{(\phi)}: \cB_\Gamma & \to & \{0,1,2,\ldots\}, \nonumber \\
  v & \to & E_\Gamma^{(\phi)}(v)
\eea
that assigns to every vertex $v$ the number of bare external $\phi$-lines
$E_\Gamma^{(\phi)}(v)$ attached to $v$.
Every such $\phi$-line represents a field $\phi$.
The number of external $\phi$-lines of $\Gamma$ is denoted by
$E_\Gamma^{(\phi)}=\sum_{v\in\cB_\Gamma} E_\Gamma^{(\phi)}(v)$.
Similarly, $E_\Gamma^{(U)}$ is a map
\bea
  E_\Gamma^{(U)}: \cM_\Gamma & \to & \{0,1,2,\ldots\}, \nonumber \\
  m & \to & E_\Gamma^{(U)}(m)
\eea
that assigns to every multiple line $m$ the number of external $U$-lines
$E_\Gamma^{(U)}(m)$ attached to $m$.
Every such $U$-line represents a field $U$ associated with a lattice link.
The number of external $U$-lines of $\Gamma$ is given by
$E_\Gamma^{(U)}=\sum_{m\in\cM_\Gamma} E_\Gamma^{(U)}(m)$.

Furthermore, $\Phi_\Gamma$ and $\Psi_\Gamma$ are incidence 
relations that assign bare internal lines
to their endpoint vertices and to their
multiple lines, respectively.
We treat lines as unoriented.
The generalization to oriented lines is easily done.
More precisely, let
$\overline{(\cB_\Gamma\times\cB_\Gamma)}^{\,\prime}$
be the set of unordered pairs of vertices $(v,w)$
with $v,w\in\cB_\Gamma$, $v\not= w$.
(The bar implies unordered pairs,
the prime the exclusions of $(v,v)$, $v\in\cB_\Gamma$.)
As in the standard linked cluster expansion, self-lines are
excluded. Every bare internal line is then mapped onto its pair of
endpoints via
\be
  \Phi_\Gamma: \cL_\Gamma \to 
   \overline{(\cB_\Gamma\times\cB_\Gamma)}^{\,\prime}.
\ee
We say that $v$ and $w$ are the endpoint vertices of $l\in\cL_\Gamma$
if $\Phi_\Gamma(l)=(v,w)$. If there is such an
$l\in\cL_\Gamma$, $v$ and $w$ are called neighbours.
Similarly, $\Psi_\Gamma$ is a map
\bea
  \Psi_\Gamma: \cL_\Gamma & \to & \cM_\Gamma, \nonumber \\
     l & \to & \Psi_\Gamma(l)
\eea
that maps every bare internal line to a multiple line.
A multiple line $m\in\cM_\Gamma$ is composed of
bare internal lines $l\in\cL_\Gamma$ which 
belong to $m$ in the sense that
$\Psi_\Gamma(l)=m$.
$l_{\cM_\Gamma}(m)$ is the total number of bare internal lines
belonging to $m$.
With $\nu=l_{\cM_\Gamma}(m)+E_\Gamma^{(U)}(m)$,
$m$ is called a $\nu$-line.
We always require that $\nu\geq 1$.
On the other hand,
every bare internal line belongs to one and only one multiple line.
For simplicity we often identify a $1$-line with the only one
bare line that belongs to it.
\vskip7pt
Next we introduce some further notions that will be used later.
External vertices are vertices having external $\phi$-lines attached,
\be
  \cB_{\Gamma,ext} =  \{ v\in\cB_\Gamma \; \vert \;
   E_\Gamma^{(\phi)}(v)\not=0 \},
\ee
whereas internal vertices do not,
$\cB_{\Gamma,int}=\cB_\Gamma\setminus\cB_{\Gamma,ext}$.
Similarly, external multiple lines have external 
$U$-lines attached,
\be
  \cM_{\Gamma,ext} =  \{ m\in\cM_\Gamma \; \vert \;
   E_\Gamma^{(U)}(m)\not=0 \},
\ee
and the complement in $\cM_\Gamma$ are the internal multiple lines,
$\cM_{\Gamma,int}=\cM_\Gamma\setminus\cM_{\Gamma,ext}$.

For every pair of vertices $v,w\in\cB_\Gamma$, $v\not= w$,
let $\overline\Phi^1(v,w)$ be the set of lines with
endpoint vertices $v$ and $w$, and
$\vert\overline\Phi^1(v,w)\vert$
the number of these lines. Thus
$\overline\Phi^1(v,w)$ is the set of lines $v$ and $w$ have
in common.
With $E_\Gamma^{(\phi)}(v)$ denoting the number of
external $\phi$-lines attached to $v\in\cB_\Gamma$,
\be
   t_{\cB_\Gamma}(v) \; = \;
   \sum_{w\in\cB_\Gamma} \vert\overline\Phi^1(v,w)\vert
    \; + \; E_\Gamma^{(\phi)}(v)
\ee
is the total number of bare lines attached to $v$.

\vskip7pt

Some topological notions and global properties of graphs will be
of major interest in the following.
A central notion is the connectivity of a \mlgraph.
Recall that we want to consider the DLCE expansion of the free energy
and of truncated correlation functions as an expansion
in connected graphs.
As indicated in section 2,
the main generalization compared to the common notion of
connectivity of a graph which is required here is that an additional
type of connectivity is provided by multiple-lines.
To define the connectivity of a \mlgraph $\Gamma$,
$\Gamma$ first is mapped to a (standard) LCE graph $\overline{\Gamma}$
to which the standard notion of connectivity applies.
There are various equivalent ways to define such a map.
We choose the following one.

\begin{itemize}

\item
For every multiple-line $m\in\cM_\Gamma$
define a new vertex $w(m)$. Let
$\widetilde\cB_\Gamma = \{ w(m) \vert m\in\cM_\Gamma \}$
and define
$\overline\cB = \cB_\Gamma \cup \widetilde\cB_\Gamma$
as the union of the vertices of $\Gamma$ and the new set of
vertices originating from the multiple-lines.

\item
For every bare internal line $l\in\cL_\Gamma$
define two new internal lines
$l_1,l_2$ and incidence relations
\bea
   \overline\Phi(l_1) & = & (v_1, w(\Psi_\Gamma(l))), \nonumber \\
   \overline\Phi(l_2) & = & (v_2, w(\Psi_\Gamma(l))),
\eea
where $v_1$ and $v_2$ are the two endpoint vertices of $l$.
The set of all lines
$l_1,l_2$, for all $l\in\cL_\Gamma$,
is denoted by $\overline\cL$.

\item
Define the external incidence relations
\bea
  \overline{E}\quad:\quad \overline\cB & \to & \{0,1,2,\ldots\}, \nonumber \\
      \overline{E}(v) & = & E_\Gamma^{(\phi)}(v),
        \quad \mbox{for $v\in\cB_\Gamma$}, \\
     \overline{E}(v) & = & E_\Gamma^{(U)}(m),
        \quad \mbox{for $v=w(m)\in\widetilde\cB_\Gamma$}.
    \nonumber
\eea

\end{itemize}

Now, $\overline\Gamma$ is defined by
\be
    \overline\Gamma \; = \; (\overline\cL,
    \overline\cB, \overline{E}, \overline\Phi ).
\ee

Having defined the standard LCE graph $\overline\Gamma$
for any \mlgraph $\Gamma$, we call
$\Gamma$ multiple-line connected or just connected 
if $\overline\Gamma$ is connected
(in the usual sense).
In Fig.~4 we have given two examples for a connected (upper graph)
and a disconnected (lower graph) \mlgraph.

\begin{figure}[h]

\begin{center}
\setlength{\unitlength}{0.8cm}

%
%
\begin{picture}(15.0,7.0)

%
%

\epsfig{bbllx=-333,bblly=50,
        bburx=947,bbury=888,
        file=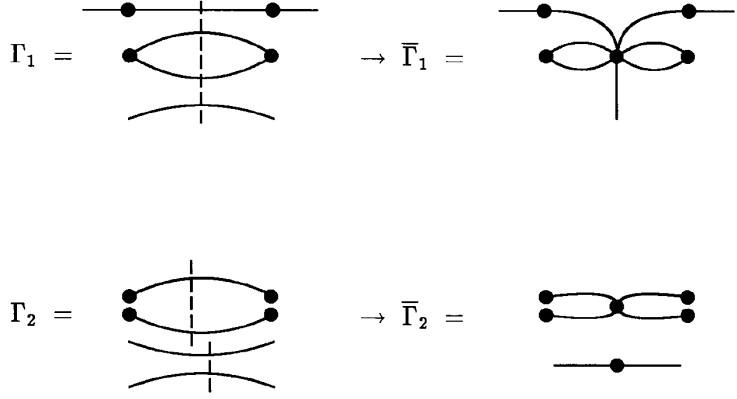,
        scale=0.22}

%
%

\end{picture}
%
%
\end{center}

\caption{\label{connected} Example of multiple-line connectivity.
The upper \mlgraph $\Gamma_1$ is connected because the graph
$\overline\Gamma_1$ is connected in the conventional sense.
The lower \mlgraph $\Gamma_2$ is disconnected because
$\overline\Gamma_2$ is so.  
}
\end{figure}

\vskip7pt

The next important notion is the topological equivalence of two
\mlgraphs.
Two \mlgraphs
\be \label{top.equiv}
  \Gamma_i \; = \; (\cL_i, \cM_i, \cB_i,
   E^{(\phi)}_i, E^{(U)}_i, \Phi_i, \Psi_i), \quad
   i=1,2
\ee
are called (topologically) equivalent if there are three
invertible maps
\bea
   \phi_1:\cB_1 &\to& \cB_2 , \nonumber \\
   \phi_2:\cL_1 &\to& \cL_2, \\
   \phi_3:\cM_1 &\to& \cM_2, \nonumber
\eea
such that
\bea
  \Phi_2 \circ \phi_2 &=& \overline\phi_1 \circ \Phi_1, \nonumber \\
  \Psi_2 \circ \phi_2 &=& \phi_3 \circ \Psi_1,
\eea
and
\bea
  E_2^{(\phi)} \circ \phi_1 &=& E_1^{(\phi)}, \nonumber \\
  E_2^{(U)} \circ \phi_3 &=& E_1^{(U)}.
\eea
Here $\circ$ means decomposition of maps, and
\bea 
  \overline\phi_1: \overline{\cB_1\times\cB_1}^{\,\prime} & \to &
      \overline{\cB_2\times\cB_2}^{\,\prime} \nonumber \\
   \overline\phi_1(v,w)  & = & (\phi_1(v),\phi_1(w)).
\eea

A symmetry of a \mlgraph
$\Gamma = (\cL,\cM,\cB,E^{(\phi)},E^{(U)},\Phi,\Psi)$
is a triple of maps
$\phi_1:\cB\to\cB$, $\phi_2:\cL\to \cL$
and $\phi_3:\cM\to\cM$
such that
\bea 
  \Phi \circ \phi_2 &=& \overline\phi_1 \circ \Phi, \nonumber \\
  \Psi \circ \phi_2 &=& \phi_3 \circ \Psi,
\eea
and
\bea
  E^{(\phi)} \circ \phi_1 &=& E^{(\phi)} \nonumber \\
  E^{(U)} \circ \phi_3 &=& E^{(U)}.
\eea
The number of these maps is called the symmetry number of $\Gamma$.

We denote by $\cG_{E_1,E_2}(L)$ the set of equivalence classes
of connected \mlgraphs with $L$ bare internal lines,
$E_1$ external $\phi$-lines and $E_2$ external $U$-lines.
Furthermore we set
\be
  \cG_{E_1,E_2} \; := \; \bigcup\limits_{L\geq 0} \; \cG_{E_1,E_2}(L).
\ee

A multiple line graph $\Gamma$ does not need to have a vertex.
If $\cB_\Gamma =0$, we have $\cL_\Gamma =0$ as well.
If in addition $\Gamma$ is connected, $\cM_\Gamma$ consists of only one element,
with all external $U$-lines attached to it.
(We anticipate that $\Gamma$ is 1-multiple-line irreducible (1MLI) by definition.
For the definition of 1MLI cf.~section 4 below.)
The only graph of $\cG_{0,E}(L=0)$ is given by
\be
  \Gamma \; = \; \left.
{
\setlength{\unitlength}{0.8cm}
%
%
\begin{picture}(4.0,1.0)

%
%
\put(0.0,0.0){
\setlength{\unitlength}{1.0cm}
\begin{picture}(10.0,0.0)

\qbezier(0.9,0.6)(1.5,0.5)(2.1,0.6)
\qbezier(0.9,0.4)(1.5,0.3)(2.1,0.4)
\qbezier(0.9,-0.5)(1.5,-0.4)(2.1,-0.5)
\qbezier(0.9,-0.3)(1.5,-0.2)(2.1,-0.3)
\put(1.5,-0.55){\line(0,1){0.15}}
\put(1.5,-0.35){\line(0,1){0.15}}
\put(1.5,-0.15){\line(0,1){0.15}}
\put(1.5,0.05){\line(0,1){0.15}}
\put(1.5,0.25){\line(0,1){0.15}}
\put(1.5,0.45){\line(0,1){0.15}}
\put(1.5,0.65){\line(0,1){0.15}}
\put(1.2,0.15){\makebox(0.2,0){$\cdot$}}
\put(1.2,-0.05){\makebox(0.2,0){$\cdot$}}
\put(1.2,0.05){\makebox(0.2,0){$\cdot$}}

\end{picture}
}
%
%

\end{picture}
%
%
}
   \right\} \; E .
\ee
It represents the leading term of the susceptibility
\be \label{susc.0E}
  \chi_{0,E} \; = \;
  \frac{1}{V D} \sum_{l_1,\dots ,l_E \in\overline\Lambda_1}
    < U(l_1) \cdots U(l_E) >^c
\ee
and is given by
$\left.{\partial^E W^1(I)}/{\partial I^E}\right\vert_{I=0}$. 
The index $c$ in (\ref{susc.0E})
stands for truncated (connected) correlation.

\vskip7pt

By removal of a $\nu$-line $m\in\cM_\Gamma$ we mean that $m$ is
dropped together with all bare internal lines and all external
$U$-lines that belong to $m$.
This notion is explained in Fig.~5a.
(It is used in section 4 for $1$-lines to define
1-particle irreducible (1PI) and 
1-line irreducible (1LI) \mlgraphs.)

On the other hand, by decomposition of a $\nu$-line $m\in\cM_\Gamma$
we mean that $m$ is dropped together with the external $U$-lines
of $m$, but all bare internal lines that belong to $m$
are kept in the graph, being identified now with $1$-lines.
This notion will be used below to define
1MLI and renormalized \mlmoments .
It is illustrated in Fig.~5b.

Similarly, decomposition of a vertex $v\in\cB_\Gamma$
means to remove the vertex $v$ and to attach the free end of every line
that entered $v$ before to a new vertex, a separate one for each line.
This notion is used to define 1-vertex-irreducible (1VI)
and renormalized vertex moments
for \mlgraphs. For an example see Fig.~5c.

\begin{figure}[h]

\begin{center}
\setlength{\unitlength}{0.8cm}

%
%
\begin{picture}(15.0,7.0)

%
%

\epsfig{bbllx=-333,bblly=60,
        bburx=947,bbury=878,
        file=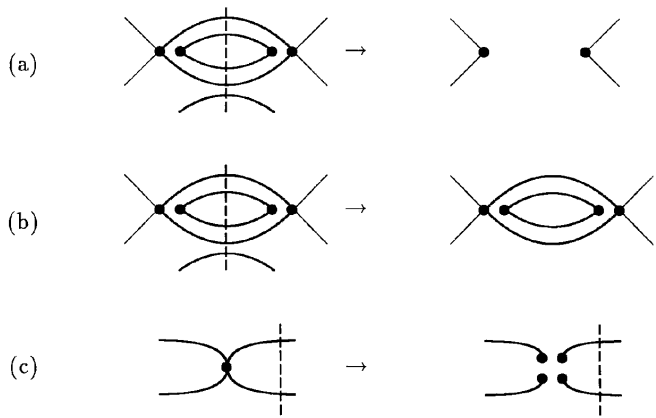,
        scale=0.20}

%
%

\end{picture}
%
%
\end{center}

\caption{\label{removdecomp} Removal (a) and decomposition (b) of
a $5$-line. Decomposition (c) of a vertex.}
\end{figure}

%
%

\subsection{Susceptibilities and weights}

In the last section
we have defined \mlgraphs and the notions of connectivity and 
equivalence of such graphs.
The definition is chosen in such a way 
that the series expansions
of the free energy and of truncated correlation functions are
obtained as a sum over equivalence classes of 
connected \mlgraphs.
The number $L$ of bare internal lines of a \mlgraph $\Gamma$
counts the order in the expansion parameter $v(x,y)$ to which
$\Gamma$ contributes.
If $v(x,y)$ is of the form
\be \label{susc.vform}
   v(x,y) \; = \; 2K \sum_{z\in\cN(x)} \delta_{y,z},
\ee
with $\cN(x)$ any finite
$x$-dependent set of lattice sites,
the contribution of $\Gamma$ is a multiple of $(2K)^L$.
Often used special cases are the nearest neighbour interactions
\be \label{susc.nn}
   v(x,y) \; = \; 2K \sum_{\mu=0}^{D-1}
   \left( \delta_{x,y+\widehat\mu} + \delta_{x,y-\widehat\mu}
   \right)
\ee
and the uniform interaction
\be \label{susc.mf}
   v(x,y) \; = \; 2K \; \left( 1 - \delta_{x,y} \right),
\ee
which is used in models of spin glasses and
partially annealed neural networks.
\vskip7pt
Susceptibilities of the $\phi$
and $U$ fields will be represented as
\bea \label{susc.susc}
  \chi_{E_1,E_2} & = &
  \frac{1}{V D} \sum_{x_1,\dots ,x_{E_1} \in \Lambda_0}
  \sum_{l_1,\dots ,l_{E_2} \in \overline\Lambda_1}
  < \phi(x_1) \cdots \phi(x_{E_1}) \;
    U(l_1) \cdots U(l_{E_2}) >^c \nonumber \\
  & \equiv & \frac{1}{V D} \sum_{x_1,\dots ,x_{E_1} \in \Lambda_0}
  \sum_{l_1,\dots ,l_{E_2} \in \overline\Lambda_1}
  \left .\frac{\partial^{E_1+E_2}W(H,I,v)}
       {\partial H(x_1)\cdots\partial H(x_{E_1})
        \partial I(l_1)\cdots\partial I(l_{E_2})}
  \right\vert_{H=I=0} \nonumber \\
   & = & \sum_{L\geq 0} (2K)^L 
  \sum_{\Gamma\in\cG_{E_1,E_2}(L)} w(\Gamma)
\eea
with lattice volume $V$ and dimension $D$.
Similar representations hold for higher moments $\mu$.

The weight $w(\Gamma)$ of a \mlgraph 
$\Gamma\in\cG_{E_1,E_2}(L)$ is given as the product of the
following factors

\begin{itemize}

\item
for every vertex $v\in\cB_\Gamma$ a factor
\be
   v^{\circ c}_n \; = \;
  \biggl( \frac{\partial^n W(H)^\circ}{\partial H^n} \biggr)_{H=0} ,
\ee
where $n=t_{\cB_\Gamma}(v)$ is the total number of bare lines
attached to $v$.

\item
for every multiple line $m\in\cM_\Gamma$ a factor
\be \label{susc.m1c}
   m^{1 c}_\nu \; = \;
 \biggl(  \frac{\partial^\nu W^1(I)}{\partial I^\nu} \biggr)_{I=0},
\ee
where $\nu=l_{\cM_\Gamma}(m)+E_\Gamma^{(U)}(m)$, that is
$m$ is a $\nu$-line,

\item
a factor $1/S_\Gamma$, where $S_\Gamma$ is the topological
symmetry number of $\Gamma$,

\item
a factor counting the permutation symmetry of external $\phi$-lines,
\be
   \frac{ E_\Gamma^{(\phi)}! }
   { \prod_{v\in\cB_\Gamma} E_\Gamma^{(\phi)}(v)! }  \quad ,
\ee

\item
a factor counting the permutation symmetry of external $U$-lines,
\be
   \frac{ E_\Gamma^{(U)}! }
   { \prod_{m\in\cM_\Gamma} E_\Gamma^{(U)}(m)! }  \quad ,
\ee

\item
the lattice embedding number of $\Gamma$, which is
the number of ways $\Gamma$ can be embedded on a lattice of
given geometry, e.~g.~on a hypercubic lattice.
To this end, the vertices of $\Gamma$ (if any) are placed onto lattice sites.
One arbitrary vertex is placed at a fixed lattice site, in order to
account for the volume factor $1/V$ in (\ref{susc.susc}).
A priori there is no exclusion principle.
This means that any number of vertices can be placed at the
same lattice site.
(This is sometimes called free embedding.)
Two restrictions apply to the embeddings.
The first constraint results from the fact that
a bare internal line represents a hopping propagator $v(x,y)$, with
lattice sites $x$ and $y$ at which the two endpoint vertices
of the line are placed at.
A reasonable computation of the embedding number takes into
account the particular form of $v(x,y)$ from the very beginning.
The second constraint is that
bare lines of the same multiple-line have to be mapped
on the same pair of sites.

For example, if $v(x,y)$ is the nearest neighbour interaction
(\ref{susc.nn}), two vertices which have 
at least one line in common are to be
placed at nearest neighbour lattice sites.
On the other hand, a propagator $v(x,y)$ of the form
(\ref{susc.mf}) implies a rather weak constraint in that
$x$ and $y$ must be different, but otherwise can be freely placed
over the lattice.

\end{itemize}

We remark that in case of a non-trivial internal symmetry
(such as considered in section 7) the expressions of
Eq.s~(\ref{susc.susc})-(\ref{susc.m1c})
must be modified appropriately. In particular,
the weight (\ref{susc.m1c}) of a multiple-line 
does no longer take such a simple form.
Eq.~(\ref{su2.Labgraph}) below is an example for the case
of a hopping term originating in an SU(2) Higgs model. 
%
%
%
%

\section{Renormalization}

Truncated correlation functions, susceptibilities
and other moments are obtained
as sums over \mlgraphs that are connected.
Their number rapidly grows with increasing order, that is with
increasing number of bare internal lines.
The procedure of "renormalization" means that the connected moments 
are represented
in terms of reduced ones. The reduced moments are obtained by summation
over \mlgraph classes which are more restricted than just by their
property of being connected.
Of course the number of graphs of such classes is smaller.
Only the most restricted \mlgraph classes must be constructed.
The subsequent steps towards the moment computation are most
conveniently done by operating analytically with the reduced
moments.
In particular, it is no longer necessary to generate all connected
and the corresponding intermediate \mlgraph classes.

A connected \mlgraph $\Gamma$ is called 1-particle irreducible (1PI) if
it satisfies the following condition.
Remove an arbitrary $1$-line of $\Gamma$. There is at most one
connected component left that has external lines attached.
(This notion is the same as the one used in the context of 
Feynman graphs.)
On the other hand, if in addition the remaining graph is still
connected, then $\Gamma$ is called 1-line irreducible (1LI).
In many cases it is sufficient to use only the second notion.
It is for instance sufficient that all vertices are constrained to have
only an even number of lines attached, or more generally, if
graphs and subgraphs with one external line are forbidden.
For notational simplicity we assume in the
following that this is the case and henceforth refer only to the notion 1LI
\footnote{In ref. \cite{LW1,thomas1} the term 1PI was used instead.}.
The generalization to the case in which 1LI and 1PI graphs must be
distinguished goes along the same lines
as for LCEs, which was discussed in \cite{thomas3}.

By $\cG_{E_1,E_2}^{1LI}(L)$ we denote the subset of \mlgraphs
$\Gamma\in\cG_{E_1,E_2}(L)$ that are 1LI.
1LI-susceptibilities are defined as series in the hopping
parameter similarly as in (\ref{susc.susc})
by restricting the summation to 1LI graphs,
\be
  \chi_{E_1,E_2}^{1LI} =
   \sum_{L\geq 0} (2K)^L 
  \sum_{\Gamma\in\cG_{E_1,E_2}^{1LI}(L)} w(\Gamma) .
\ee
Susceptibilities are easily obtained in a closed form in terms of
1LI-susceptibilities $\chi^{1LI}$. It can be shown that the $\chi^{1LI}$s
can be obtained by an appropriate Legendre transform.
For instance
\bea
  \chi_{2,0} & = & \frac{\chi_{2,0}^{1LI}}
    {1 - \widetilde{v}(0) \chi_{2,0}^{1LI}} , \nonumber \\
   \chi_{2,1} & = & \frac{\chi_{2,1}^{1LI}}
    {(1 - \widetilde{v}(0) \chi_{2,0}^{1LI})^2} ,
\eea
where $\widetilde{v}(k)$ is the Fourier transform of the hopping
propagator $v(x,y)$,
\be
  v(x,y) \; = \; \int_{-\pi}^\pi \frac{d^Dk}{(2\pi)^D} \;
   e^{-ik\cdot (x-y)} \; \widetilde{v}(k) .
\ee

In LCEs the second important resummation comes from so called
vertex renormalizations.
This means partial resummation of graphs with specific properties
such as having one external vertex only. These sums then are
considered as "renormalized vertices" replacing the vertices
of graphs with complementary properties.
The procedure naturally leads to the notion of 1-vertex irreducibility
(1VI) and renormalized moments.

In DLCE we follow this procedure. The very definition of 1VI has
to be modified slightly for \mlgraphs because of the enhanced connectivity
properties due to multiple-lines.
In addition, as a natural generalization,
we supplement vertex renormalization by multiple-line renormalization.

A \mlgraph $\Gamma$ is called 1-vertex irreducible (1VI) if it
satisfies the following condition.
Decompose an arbitrary vertex $v\in\cB_\Gamma$.
Every connected component of the remaining graph has then at least one
external line attached. It can be a $\phi$-line or a $U$-line.
We write
\be
   \cG_{E_1,E_2}^{1VI}(L) \; = \;
   \{ \Gamma\in\cG_{E_1,E_2}^{1LI}(L) \; \vert \;
     \mbox{$\Gamma$ is 1VI} \}
\ee
for the set of equivalence classes of graphs that are both 1LI and 1VI,
with $E_1$ external $\phi$-lines, $E_2$ external $U$-lines
and $L$ bare internal lines.

The renormalized vertex moment graphs are 1LI graphs that have
precisely one external vertex and no external multiple line,
\be
  Q_k(L) \; = \;
  \{ \Gamma\in\cG_{k,0}^{1LI}(L) \; \vert \;
     \mbox{there is $v\in\cB_\Gamma$ with $E_\Gamma^{(\phi)}(v)=k$} \}.
\ee

A \mlgraph $\Gamma$ is called 1-multiple-line irreducible (1MLI) if
it satisfies the following criterion.
Decompose an arbitrary multiple-line $m\in\cM_\Gamma$.
Every remaining connected component has then at least one external line
attached. It can be a $\phi$-line or a $U$-line.
We write
\be
   \cG_{E_1,E_2}^{1MLI}(L) \; = \;
   \{ \Gamma\in\cG_{E_1,E_2}^{1LI}(L) \; \vert \;
     \mbox{$\Gamma$ is 1MLI} \}.
\ee
The renormalized multiple-line moment graphs are graphs that are
1LI and have precisely one external multiple-line, but no external
vertex,
\be
  R_k(L) \; = \;
  \{ \Gamma\in\cG_{0,k}^{1LI}(L) \; \vert \;
     \mbox{there is $m\in\cM_\Gamma$ with $E_\Gamma^{(U)}(m)=k$} \}.
\ee

The equivalence classes of graphs that are both 1VI and 1MLI
are denoted by
\be
   S_{E_1,E_2}(L) \; = \;
      \cG_{E_1,E_2}^{1VI}(L) \cap \cG_{E_1,E_2}^{1MLI}(L).
\ee

With the renormalized moment graphs as defined above, the
1LI-susceptibilities are now obtained in the form
\be
  \chi_{E_1,E_2}^{1LI} =
   \sum_{L\geq 0} (2K)^L 
  \sum_{\Gamma\in\cS_{E_1,E_2}(L)} \widetilde{w}(\Gamma) .
\ee
The weights $\widetilde{w}(\Gamma)$ are given as a product
of factors as described in the last subsection, with
the following two exceptions.

\begin{itemize}

\item
The vertex coupling constants $v^{\circ c}_n$ are replaced
by the renormalized vertex moments
\be
  v^{\circ c}_n \; \to \;
  v^c_n = \sum_{L\geq 0} (2K)^L
  \sum_{\Gamma\in Q_n(L)} w(\Gamma) .
\ee

\item
The multiple line coupling constants $m^{1 c}_\nu$ are replaced
by the renormalized multiple line moments
\be
  m^{1 c}_\nu \; \to \;
  m^c_\nu = \sum_{L\geq 0} (2K)^L
  \sum_{\Gamma\in R_\nu (L)} w(\Gamma) .
\ee

\end{itemize}

In the series representations above, the $w(\Gamma)$
are computed according to the rules
of subsection 3.2.
%
%
%
%

\section{Graph construction}

In this section we describe a mechanism to generate the
\mlgraph classes
$\cS_{E_1,E_2}(L)$, $\cQ_k(L)$ and $\cR_k(L)$
with $L\geq 1$ internal lines.
Similar as in LCEs the idea is to define appropriate classes of vacuum 
graphs
(no external lines). The generation of the graphs of
$\cS$, $\cQ$ and $\cR$ is then done by attaching external lines in 
various ways.
By defining appropriate recursion relations, the order by order
construction can then be restricted to the vacuum classes.

Fortunately we can profit from graphs that have 
been generated already in standard LCEs.
This way we completely avoid a cumbersome recursive construction, but
obtain the vacuum \mlgraphs by operating on the
LCE vacuum graphs of the same order.

The graphs of $\cS$, $\cQ$ and $\cR$ will be obtained through a
sequence of simpler graph classes.
At each step,
any method of \mlgraph construction should satisfy the following two
conditions.

\begin{itemize}

\item
It should be surjective or complete. At least one graph
of every equivalence class should be generated.

\item
For every equivalence class precisely one representant should be
kept. All generated graphs of a particular class of graphs
should be mutually inequivalent.
This requires a so called
non-equivalence test for every newly generated graph. The graph should
be kept if and only if it is not equivalent to a graph
already generated.
In general it must be compared with all \mlgraphs that have been 
generated before. If this is necessary, the non-equivalence test
is rather time consuming.

\end{itemize}

The algorithm we describe below circumvents an extensive
comparison of \mlgraphs .
It is defined in such a way that two graphs can be equivalent
only if they have their origin in the same graph of the prior
graph class.

Essentially we proceed in the following steps.
The starting point are the vacuum diagrams of $\cP_2(L)$
as computed in \cite{thomas1}.
$\cP_2(L)$ is the set of (standard) LCE equivalence classes of
connected, 1LI vacuum graphs with $L$ lines.
Operating on $\cP_2(L)$, we generate all mutually inequivalent
\mlgraphs with $L$ lines that are connected, 1LI and have no
external lines (neither $\phi$-lines nor $U$-lines),
\be
   \cM\cP_2(L) \; \equiv \; \cG_{0,0}^{1LI}(L).
\ee
This is done in two steps.
In a first step lines with the same endpoint vertices
are arranged in all possible ways into multiple lines.
We obtain a subset of $\cM\cP_2(L)$ which we denote by
$\widetilde{\cM\cP_2}(L)$.
A \mlgraph $\Gamma\in\cM\cP_2(L)$ belongs to
$\widetilde{\cM\cP_2}(L)$,
$\Gamma\in\widetilde{\cM\cP_2}(L)$,
if and only if all bare lines of $\Gamma$ that belong to the same
multiple line have the same endpoint vertices.
The second step is to relax the additional constraint on
$\widetilde{\cM\cP_2}(L)$.
This is achieved by resolving the vertices.
This means that every vertex is split into several ones, and
that the bare lines that were attached to the original vertex
before now are distributed over the new vertices
in all possible ways.
As a general constraint on the resolutions, the resulting
\mlgraphs should be 1LI in order to stay in
$\cM\cP_2(L)$.
We impose an additional constraint on the resolutions, allowing
only for so-called admissible vertex resolutions.
They are defined in such a way that two \mlgraphs of
${\cM\cP_2}(L)$
can be equivalent only if they originate from the same graph
of $\widetilde{\cM\cP_2}(L)$.
This considerably simplifies the non-equivalence test.

Once we have obtained ${\cM\cP_2}(L)$,
the final \mlgraph classes
$\cS_{E_1,E_2}(L)$, $\cQ_k(L)$ and $\cR_k(L)$
are obtained by attaching external $\phi$-lines and $U$-lines
in the appropriate way to all graphs
of ${\cM\cP_2}(L)$.
The steps in the generation of the DLCE graph classes
are shown schematically in Fig.~6.

\begin{figure}[h]

\begin{center}
\setlength{\unitlength}{0.8cm}

%
%
\begin{picture}(15.0,2.5)

\put(0.0,1.0){\makebox(2.0,0){$\cP_2(L)$}}
\put(2.3,1.0){\vector(1,0){0.9}}
\put(4.0,1.0){\makebox(2.0,0){$\widetilde{\cM\cP}_2(L)$}}
\put(6.3,1.0){\vector(1,0){0.9}}
\put(8.0,1.0){\makebox(2.0,0){$\cM\cP_2(L)$}}
\put(10.3,1.0){\vector(1,1){0.9}}
\put(10.3,1.0){\vector(1,0){0.9}}
\put(10.3,1.0){\vector(1,-1){0.9}}
\put(12.0,2.0){\makebox(2.0,0){$\cS_{E_1,E_2}(L)$}}
\put(12.0,1.0){\makebox(2.0,0){$\cQ_k(L)$}}
\put(12.0,0.0){\makebox(2.0,0){$\cR_k(L)$}}

\end{picture}
%
%
\end{center}

\caption{\label{history} Modules in the generation of DLCE graph classes.
The root of the inheritance tree are the LCE vacuum graphs of
$\cP_2(L)$.
By means of multiple-line construction
($\widetilde{\cM\cP}_2$) and vertex resolution
($\cM\cP_2$) the DLCE
vacuum graphs of $\cM\cP_2(L)$ are obtained.
Attaching external $\phi$-lines and external $U$-lines in well defined
ways yields the renormalized vertices ($\cQ_k(L)$),
the multiple-line moments ($\cR_k(L)$), and
the 1MLI and 1VI graphs ($\cS_{E_1,E_2}(L)$).}
\end{figure}
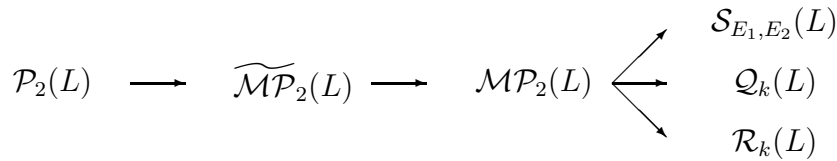

In the following we describe the \mlgraph construction in detail.
Below we use the following notation.
For any finite set $\cA$ let $\cP(\cA)$ denote the set of partitions
of $\cA$ into mutually disjoint nonempty subsets of $\cA$.
For every $\Pi\in\cP(\cA)$ we have
\be
   \cA \; = \; \bigcup_{P\in\Pi} P.
\ee

%
%

\subsection{$\cP_2(L) \to \widetilde{\cM\cP}_2(L)$}

Starting with the LCE vacuum graphs of $\cP_2(L)$,
we first of all arrange internal lines into multiple lines.
Consider an arbitrary graph
$\Gamma\in\cP_2(L)$.
Simultaneously for every pair of neighbouring 
vertices $v,w\in\cB_\Gamma$ choose a partition
$\Pi\in\cP(\cL_{v,w})$,
where $\cL_{v,w}$ is the set of common lines between 
the vertices $v$ and $w$.
Every $l\in\cL_{v,w}$ belongs to a unique $P\in\Pi\in\cP(\cL_{v,w})$.
For every $P\in\Pi$ we introduce a multiple line $m(P)$ such that
all $l\in P$ are precisely the bare internal lines that belong
to $m(P)$.
This procedure defines the incidence relation $\Psi(l)$ for all
$l\in\cL_\Gamma$
and, along with that, a \mlgraph that belongs to
$\widetilde{\cM\cP_2}(L)$.

In Fig.~7 we give examples for this construction.
In the first one, an LCE graph leads to 5 mutually inequivalent
DLCE gaphs,
in the second one to 3 DLCE graphs of $\widetilde{\cM\cP}_2(L)$.

Proceeding this way for all partitions of lines for all pairs of neighboured
vertices of $\Gamma$,
and for all $\Gamma\in\cP_2(L)$,
we obtain the class of \mlgraphs
$\widetilde{\cM\cP_2}(L)$.
The non-equivalence check for every newly generated \mlgraph
$\widetilde\Gamma$ can be drastically restricted.
It is sufficient to compare $\widetilde\Gamma$ only with those
\mlgraphs that have been constructed before from the same graph
$\Gamma\in\cP_2(L)$ as $\widetilde\Gamma$ results from.
Other graphs of $\widetilde{\cM\cP_2}(L)$ are inequivalent to
$\widetilde\Gamma$.

\begin{figure}[h]

\begin{center}
\setlength{\unitlength}{0.8cm}

%
%
\begin{picture}(15.0,7.0)

%
%

\epsfig{bbllx=-333,bblly=149,
        bburx=947,bbury=786,
        file=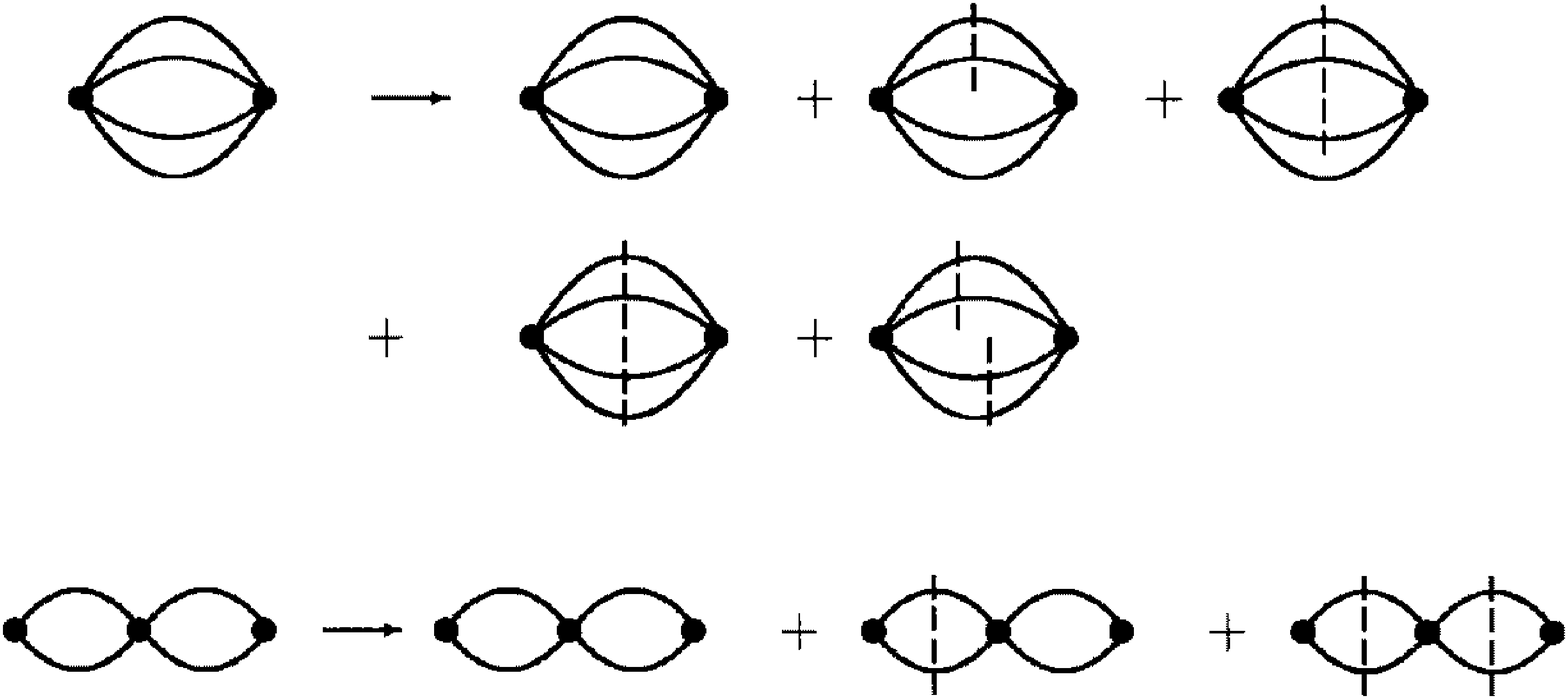,
        scale=0.25}

%
%

\end{picture}
%
%
\end{center}

\caption{\label{multline} Two examples of multiple-line construction
$\cP_2(L)$ $\to$ $\widetilde{\cM\cP}_2(L)$ with $L=4$.
Common lines between two vertices are arranged into multiple
lines in all possible ways.
Only mutually inequivalent graphs are shown.}
\end{figure}

%
%

\subsection{$\widetilde{\cM\cP}_2(L) \to \cM\cP_2(L)$}

The second step is to relax the constraint on $\widetilde{\cM\cP}_2$
that bare lines which belong to the same multiple line
must have the same endpoint vertices.
This is achieved by vertex resolutions
of all graphs of $\widetilde{\cM\cP_2}(L)$.
To efficiently reduce the non-equivalence check of new graphs
we allow only for
so-called admissible vertex resolutions.

Let $\widetilde\Gamma\in\widetilde{\cM\cP_2}(L)$,
$v\in\cB_{\widetilde\Gamma}$ any vertex of $\widetilde\Gamma$ and
$\Pi\in\cP(\cL_v)$ any partition of the set of lines
$\cL_v$ entering $v$.
We remove the vertex $v$ and draw for every $P\in\Pi$ a new
vertex $v(P)$ so that all lines $l\in P$ enter the vertex
$v(P)$ rather than $v$ before its removal.
This procedure is called a vertex resolution of $v$. 
The resulting graph $\Gamma$ must be 1LI in order to belong
to $\cM\cP_2(L)$.

The vertex resolution of $v$ is called admissible if in addition the
(standard LCE) graph $\widehat\Gamma$ 
is connected with $\widehat\Gamma$ defined in the following way.
For every $l\in\cL_v$ there is precisely one $m\in\cM_{\widetilde\Gamma}$
with $\Psi_{\widetilde\Gamma}(l)=m$,
and there is exactly one $P\in\Pi$ with $l\in P$.
Draw a vertex $w(m) = w(\Psi_{\widetilde\Gamma}(l))$
for every such multiple line,
a vertex $v(P)$ for every $P\in\Pi$,
and define a line $\widehat{l}(l)$ with incidence relation
\be
   \widehat\Phi(\widehat{l}) \; = \;
   ( w(\Psi_{\widetilde\Gamma}(l)), v(P) ).
\ee
This defines a graph $\widehat\Gamma = ( \cB_{\widehat\Gamma},
\cL_{\widehat\Gamma}, E_{\widehat\Gamma}=0, \widehat\Phi )$ with
\bea
   \cB_{\widehat\Gamma} & = &
    \{ v(P) \vert P\in\Pi \}
    \cup \{ w(\Psi_{\widetilde\Gamma}(l))
        \vert l\in\cL_v \} \nonumber \\
   \cL_{\widehat\Gamma} & = &
    \{ \widehat{l}(l) \vert l\in\cL_v \} .
\eea
A vertex resolution is called admissible if the graph
$\widehat\Gamma$ is connected.
Roughly speaking this implies that no vertex drops off
after vertex resolution.

Fig.~8 shows two admissable and one non-admissible vertex
resolutions (the auxiliary graphs $\widehat\Gamma$ are
not displayed).

\begin{figure}[h]

\begin{center}
\setlength{\unitlength}{0.8cm}

%
%
\begin{picture}(15.0,7.0)

%
%

\epsfig{bbllx=-341,bblly=76,
        bburx=667,bbury=859,
        file=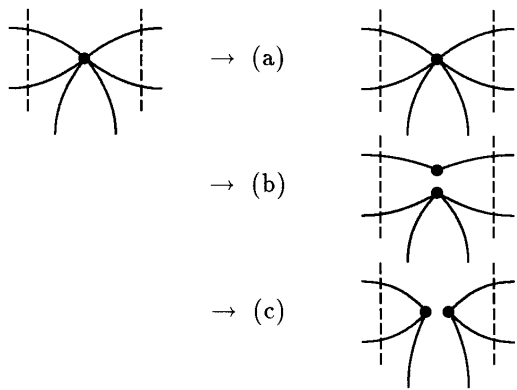,
        scale=0.2}

%
%

\end{picture}
%
%
\end{center}

\caption{\label{removal} Example of admissible ((a) and (b)) and
non-admissible (c) vertex resolutions.
In particular, the case of the trivial resolution (a)
is admissible.}
\end{figure}

The class $\cM\cP_2(L)$ is now generated as follows.
Consider an arbitrary graph 
$\widetilde\Gamma\in\widetilde{\cM\cP}_2(L)$.
Simultaneously for all vertices $v\in\cB_{\widetilde\Gamma}$ apply an
admissible vertex resolution.
We obtain a \mlgraph $\Gamma\in\cM\cP_2(L)$.
Doing this for all possible resolutions of all vertices
and for all $\widetilde\Gamma\in\widetilde{\cM\cP}_2(L)$,
we get the complete graph class
$\cM\cP_2(L)$.
Again, due to the restriction to admissible vertex resolutions,
the non-equivalence check of a newly generated graph
$\Gamma\in\cM\cP_2(L)$ can be restricted to graphs that have
been generated
from the same graph of $\widetilde{\cM\cP_2}(L)$ as $\Gamma$ was.

In Fig.~9 we show an example for the application of admissible
vertex resolutions. Here one graph of $\widetilde{\cM\cP}_2(L)$
leads to eight graphs of $\cM\cP_2(L)$, indicating the proliferation
of DLCE graphs. Note that the starting graph was only one of
five of Fig.~7.
\begin{figure}[h]

\begin{center}
\setlength{\unitlength}{0.8cm}

%
%
\begin{picture}(15.0,5.5)

%
%

\epsfig{bbllx=-333,bblly=197,
        bburx=947,bbury=738,
        file=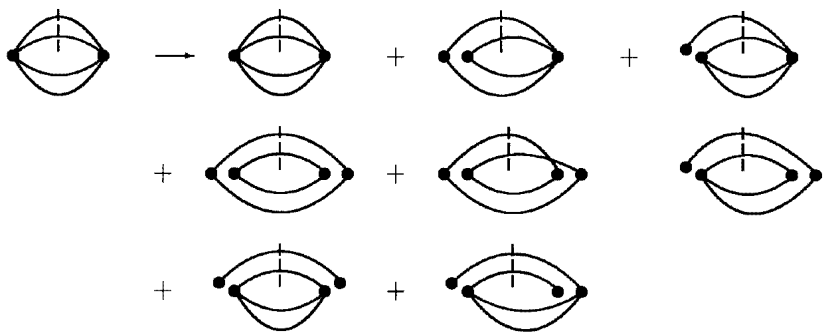,
        scale=0.25}

%
%

\end{picture}
%
%
\end{center}

\caption{\label{resolution} Example of applying admissible vertex
resolutions for $\widetilde{\cM\cP}_2(L)$ $\to$
${\cM\cP}_2(L)$ with $L=4$.
Only mutually inequivalent graphs are shown.}

\end{figure}

%
%

\subsection{$\cM\cP_2(L) \to \cS_{E_1,E_2}(L)$, $Q_k(L)$ and $R_k(L)$}

The final step is to construct the classes of \mlgraphs
$\cS_{E_1,E_2}$, $Q_k$ and $R_k$
out of the vacuum \mlgraphs of $\cM\cP_2$.
This step is realized by attaching external 
$\phi$-lines and external $U$-lines
in different ways to every \mlgraph $\Gamma\in\cM\cP_2(L)$.

$R_k(L)$ is obtained by attaching $k$ external $U$-lines
to just one multiple line $m\in\cM_\Gamma$, for all multiple lines $m$
and for all $\Gamma$.

$Q_k(L)$ is obtained by attaching $k$ external $\phi$-lines
to just one vertex $v\in\cB_\Gamma$, for all vertices $v$
and for all $\Gamma$.

$\cS_{E_1,E_2}(L)$ is obtained by attaching $E_1$ external $\phi$-lines
to the vertices of $\Gamma$ and $E_2$ external $U$-lines to the
multiple lines of $\Gamma$, for all $\Gamma\in\cM\cP_2(L)$.
This is done under the constraint that the resulting \mlgraphs
have to be 1VI and 1MLI.

Again, the non-equivalence test of \mlgraphs can be confined
to pairs of graphs that have their origin in the same graph
of $\cM\cP_2(L)$.

In Fig.~10 we illustrate the attachment of external lines to a vacuum
DLCE graph of $\cM\cP_2(L)$.

\begin{figure}[h]

\begin{center}
\setlength{\unitlength}{0.8cm}

%
%
\begin{picture}(15.0,5.5)

%
%

\epsfig{bbllx=-609,bblly=111,
        bburx=503,bbury=825,
        file=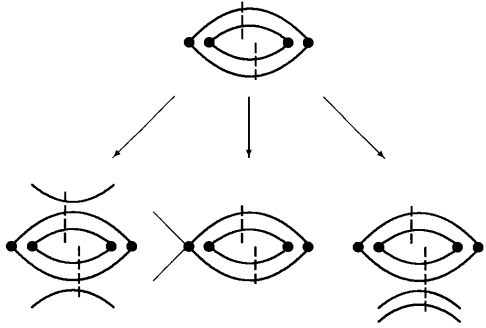,
        scale=0.18}

%
%

\end{picture}
%
%
\end{center}

\caption{\label{sqr} Attaching external lines to a vacuum DLCE graph
of ${\cM\cP}_2(L)$. The example shows from left to right:
the generation of a graph of $\cS_{0,2}(L)$, $Q_2(L)$ and $R_2(L)$,
with $L=4$.}

\end{figure}
%
%
%
%
\section{Applications to Spin Glasses and Neural Networks}

In this section we indicate applications of DLCEs to spin glasses and
partially annealed neural networks with "slow" interactions and "fast"
spins \cite{sherrington}-\cite{penney}, \cite{fischer}.

%
\subsection{The SK model and the replica trick}

For simplicity we consider the Sherrington-Kirkpatrick (SK) model
which is a spin glass model with infinite connectivity
\cite{sherrington}.
Following the notation of section 2,
the fast spins are represented by the $\Phi$-field
$\Phi_i$, $i=1,\dots ,N$, taking values in $\{-1,+1\}$.
$N$ denotes the total number of spins.
The slow interactions are mediated by the $U$-field
$U_{(i,j)}$, $i<j=1,\dots ,N$.
The SK model is described by the partition function 
\be\label{neur.z}
  Z_{\beta^\prime}^{\prime} \; = \; \cN \,
  \int_{-\infty}^\infty \prod_{i<j = 1}^N dU_{(i,j)} \;
  \exp{(-\beta^\prime \cH(U))},
\ee
with $\cN$ some normalization that will be specified below.
The effective Hamiltonian $\cH$ of
$U$ is given by
\be\label{neur.h}
 \cH(U) \; = \; - \frac{1}{\beta} \ln{Z_\beta(U)}
 + \frac{1}{2}\mu N \sum_{i<j=1}^N U_{(i,j)}^2,
\ee
where $\mu$ is a positive coupling constant.
$Z_\beta(U)$ is the partition function of a
spin system
\be\label{neur.zspin}
  Z_\beta(U) \; = \; \sum_{\{ \Phi_i = \pm 1\}}
  \exp{(\beta\sum_{i<j} U_{(i,j)} \Phi_i \Phi_j })
\ee
at frozen spin-spin couplings $U_{(i,j)}$ and temperature
$\beta^{-1}$, controlling
the fast spin fluctuations. In (\ref{neur.z}) it has been assumed that the
equilibrium distribution of the slow variables is a Boltzmann distribution
governed by a second temperature $\beta^{\prime -1}$. This assumption is
justified, if the time evolution of the $U$s is determined by a 
dissipative Langevin
equation (for the precise conditions see e.g.~\cite{zinn-justin}). 
Such a Langevin equation can be actually derived for the $U$s
from an ansatz which is motivated by neural networks
\cite{sherrington}-\cite{penney}.

One of the major quantities of interest
is $\ln Z^{\prime}_{\beta^\prime}$.
Let us first rewrite $Z^\prime_{\beta^\prime}$ in the form
\bea\label{neur.zre}
  Z_{n\beta}^{\prime} & = &
  \int_{-\infty}^\infty \prod_{i<j = 1}^N 
  \left( \sqrt{\frac{JN}{2\pi}} dU_{(i,j)} \right) \cdot
  \exp{(-\frac{1}{2} J N \sum_{i<j} U_{(i,j)}^2})
  \;\; Z_\beta(U)^n \nonumber \\
  & \equiv & [[ Z_\beta(U)^n ]] ,
\eea
where we have introduced $J=\beta\mu$ and
$n=\beta^\prime/\beta$.
The normalization has been chosen such that 
$[[1]] = 1$.

The limiting case $n=0$ with $J$ kept fixed corresponds to a
quenching of the $U$s in a spin
glass, in this limit the $U$ degrees of freedom decouple from the spin
dynamics, while $n=1$ corresponds to a pure annealing in the disordered
system (see e.g.~\cite{dotsenko}). 
Partial annealing then refers to a value of $n$ with $0<n<1$.
In particular the limit $n \rightarrow 0$ with fixed $J$ and $\beta$
amounts to the replica trick in spin glasses \cite{zinn-justin}.
Assuming that averaging $[[\cdot ]]$ and 
limes $n \rightarrow 0$ commute, we have
\be\label{neur.rep}
   [[ \ln{Z_\beta}(U) ]] \; = \;
 \lim_{n\to 0} \frac{ \ln{Z_{n\beta}^\prime }}{n}.
\ee
The left hand side of (\ref{neur.rep}) is calculated
by means of the right hand side.

\vskip7pt

So far $n$ has been a positive real number. For integer
$n=1,2,\dots$,
$Z_\beta(U)^n$ is the partition function of an n-times replicated
system.
Let us first consider this case. We rewrite
\be\label{neur.zspin1}
  Z_\beta(U)^n \; = \; \sum_{\{\Phi_i^{(a)}\}}
  \exp{( \beta \sum_{a=1}^n \sum_{i<j=1}^N U_{(i,j)}
  \Phi_i^{(a)} \Phi_j^{(a)} )} ,
\ee
with $a=1,\dots , n$ labelling the replicated spin variables, so that
\bea\label{neur.zspin2}
  Z_{n\beta}^{\prime} & = &
  \int_{-\infty}^\infty \prod_{i<j = 1}^N dU_{(i,j)} \cdot
  \sum_{\{\Phi_i^{(a)} = \pm 1\}} 
  \exp{(-S(U,\Phi^{(a)}))} , \nonumber \\
  S(U,\Phi^{(a)}) & = & 
  -\beta \sum_{a=1}^n \sum_{i<j=1}^N U_{(i,j)} \Phi_i^{(a)} \Phi_j^{(a)}
  + \frac{1}{2} J N \sum_{i<j} U_{(i,j)}^2.
\eea
Linear terms in $\Phi$ and $U$ may be included according to
\be\label{neur.zspin3}
  S_{lin} \; = \; - h \sum_{a=1}^n \sum_{i=1}^N \Phi_i^{(a)}
  + c \sum_{i<j=1}^N U_{(i,j)} 
\ee
with constant external fields $h$ and $c$.
Apparently $Z_{n\beta}^\prime$ has the form of models 
to which DLCE applies, with a hopping term 
\be\label{neur.hop}
  S_{hop}(U,\Phi^{(a)}) \; = \; - \beta \sum_{a=1}^n \sum_{i<j=1}^N
  U_{(i,j)} \Phi_i^{(a)} \Phi_j^{(a)},
\ee
a single link action
\be\label{neur.link}
 S^1(U_{(i,j)}) \; = \; c U_{(i,j)}
 + \frac{1}{2} J N \; U_{(i,j)}^2,
\ee
and a single site action
\be\label{neur.site}
 S^\circ(\Phi_i^{(a)}) \; = \;
 - h \sum_{a=1}^n \Phi_i^{(a)}.
\ee
Application of DLCE to
$\ln Z^\prime_{\beta^\prime}$ and derived quantities
yields their series expansions in $\beta$.

So far, merely for simplicity,
we have considered the SK model which is a spin glass model
with infinite and uniform connectivity, in which all spins are coupled with
equal variance
$[[ U_{(i,j)}^2]] = 1/(J N)$.
For integer $n=1,2\dots$, this system
becomes exactly solvable in the thermodynamic limit corresponding to
the limit $N \rightarrow \infty$, so that the leading order of a large-$N$
expansion becomes exact, cf.~e.g.~\cite{fischer}.
In DLCE this amounts to a resummation of appropriate tree graphs.

In cases of finite or non-uniform connectivity
it is no longer possible to obtain solutions in a closed form.
Examples are multilayered neural networks with
couplings between adjacent layers,
and spin glasses with short range interactions.
Models of this kind are suited for an application of DLCEs
and will be studied in future work.

Finally we come to the limiting case of 
$n=0$. Usually it is obtained by extrapolating the results obtained
for $n=1,2,\dots$ towards $n=0$.
For instance,
$[[\ln Z_\beta(U)]]$ is obtained from
$[[Z_\beta(U)^n]]$
for $n=1,2,\dots$.
This is a rather subtle point because the extrapolation is not unique
without further assumptions.
In the next subsection we show how it is possible to completely avoid
both the extrapolation and the replica trick and to directly compute
the quantity of interest by means of dynamical linked cluster expansions.
The task is to appropriately identify 
the classes of contributing graphs.

%
\subsection{Avoiding the replica trick}

First
we adapt the notation to section 2 to include more general cases.
$\Lambda_0$ denotes the support of the spins, 
that is the set of lattice sites, with
$V=\vert\Lambda_0\vert$ denoting their total number.
$\overline\Lambda_1$ $\subseteq\Lambda_1$ are the pairs of sites
whose spins interact.
In accordance with (\ref{neur.zre}), we write for
the normalized link-average of a function $f(U)$
\be \label{sg.average}
   [[ f(U) ]] \; = \; \int \cD U \; f(U)
\ee
with
\bea
   \cD U & = & \prod_{l\in\overline\Lambda_1}
     d\mu(U(l)) \; ,
   \nonumber \\
   d\mu(U) & = & \cN_1 \; dU \; \exp{(-S^1(U))} \; , \;
     \int_{-\infty}^\infty d\mu(U) = 1 .
\eea
It is convenient to introduce the single link expectation values
\be
  < g(U) >_1 \; \equiv \;
  \int d\mu (U) \; g(U)
\ee
and the generating function $W^1(I)$ by
\be
   \exp{W^1(I)} \; \equiv \;
   < \exp{(IU)} >_1 .
\ee

The way in which the replica trick can be avoided
is examplified for the free energy density 
$W_{sp}/V$ of the spin system
averaged over the link couplings.
The partition function of the spin system
for a given distribution of the link interactions $U(x,y)$
is given by
\be
  \exp{W_{sp}(U)} \; = \;
  \cN_{sp} \; \int \cD\Phi \;
  \exp{(-S_{sp}(\Phi,U))},
\ee
where $W_{sp}(0)=0$ and
\bea
  S_{sp}(\Phi,U) & = &
  - \; \frac{1}{2} \sum_{x,y\in\Lambda_0} v(x,y)
   \Phi(x) \Phi(y) U(x,y), 
   \nonumber \\
  \cD\Phi & = & \prod_{x\in\Lambda_0}
   d\Phi(x) \cdot \exp{(-S^\circ(\Phi(x)))}.
\eea
Without loss of generality we
identify the support of the interaction
$v(x,y)=v(y,x)$ with the set 
$\overline{\Lambda}_1$ of lattice sites where 
$\cD U$ is supported,
\be
   \overline\Lambda_1 \; = \;
   \{ l=(x,y)\in\overline{\Lambda_0\times\Lambda_0} \; \vert \;
   v(x,y)\not= 0\}.
\ee
For simplicity we assume $v(x,y)$ to be of the form
(\ref{susc.vform}),
so that $K$ is a measure of the strength of the interactions
$v(x,y)$.

\vskip7pt

The free energy density of the spin system allows for a
series expansion in the standard LCE sense, 
with the link field $U(l)$ playing the role
of a "background field",
\be \label{sg.wsp_series}
   \frac{1}{V} W_{sp}(U) \; = \;
   \sum_{L\geq0} (2K)^L 
   \sum_{\Gamma\in\cG^{sp}_0(L)}w^{sp}(\Gamma,U).
\ee
Here $\cG_E^{sp}(L)$ (with $E=0$) denotes the
set of equivalence classes of connected LCE graphs
with $E$ external lines and $L$ internal lines.
The spin-weights $w^{sp}(\Gamma,U)$ are of the form
\be \label{sg.wsp_weight}
 w^{sp}(\Gamma,U) \; = \; R^{sp}(\Gamma)
  \sum_{\cL_\Gamma\to\overline{\Lambda}_1}^{\quad\prime}
  \prod_{l\in\overline{\Lambda}_1}
  U(l)^{m(l)} .
\ee
The sum is taken over all non-vanishing lattice embeddings
of the graph $\Gamma$.
It runs over all maps of internal lines of the graph
$\Gamma$ to pairs of lattice sites of $\overline\Lambda_1$
that are consistent with the graph topology in the
sense discussed in section 3.
For every $l\in\overline\Lambda_1$, $m(l)$ denotes the number
of lines of $\Gamma$ that are mapped onto the link $l$
by the embedding.
All other factors that contribute to the weight
are collected in the prefactor $R^{sp}(\Gamma)$,
including the inverse topological symmetry number of $\Gamma$.

\vskip7pt

Next we want to express
$[[W_{sp}(U)]]$ as a series in $K$ by means of DLCE.
Toward this end we set $f(U)=W_{sp}(U)$ and insert the series
(\ref{sg.wsp_series}) with (\ref{sg.wsp_weight})
into (\ref{sg.average}).
At this stage we are not concerned with question of 
(uniform or dominated)
convergence and obtain
\bea \label{sg.wspint}
  [[ \frac{1}{V} \; W_{sp}(U) ]] & = &
   \sum_{L\geq 0} (2K)^L \sum_{\Gamma\in\cG_0^{sp}(L)}
    \int\cD U \; w^{sp}(\Gamma,U)
   \nonumber  \\
  & = & \sum_{L\geq 0} (2K)^L \sum_{\Gamma\in\cG_0^{sp}(L)}
   R^{sp}(\Gamma)  \sum_{\cL_\Gamma\to\overline{\Lambda}_1}^{\quad\prime}
  \prod_{l\in\overline{\Lambda}_1}
  <U(l)^{m(l)}>_1 .
\eea

The next step is to express the full single link expectation values
in terms of the connected ones. They are related by
\be \label{sg.conn}
   < U^m >_1 \; = \;
   \sum_{\Pi\in\cP(\underline{m})} \prod_{P\in\Pi}
   < U^{|P|} >_1^c,
\ee
where $\cP(\underline{m})$ denotes the set of all partitions
of $\underline{m}=\{1,\dots ,m\}$ into non-empty,
mutually disjoint subsets of
$\underline{m}$. 
$|P|$ is the number of elements of the set $P$.
The relation (\ref{sg.conn})
is equivalent to the partition of all lines of $\Gamma$ that are
mapped to the same lattice link into multiple lines,
with every multiple line contributing a factor
\be \label{sg.wdiff}
   < U^{|P|} >_1^c \; = \;
   \left. \frac{\partial^{|P|} W^1(I)}{\partial I^{|P|}}
   \right\vert_{I=0} \; = \;
   m^{1 c}_{|P|}.
\ee
Using (\ref{sg.conn}), (\ref{sg.wdiff}) we rewrite
(\ref{sg.wspint}) as
\be \label{sg.wsptodlce}
  [[ \frac{1}{V} \; W_{sp}(U) ]]  =
   \sum_{L\geq 0} (2K)^L \sum_{\Gamma\in\cG_0^{sp}(L)}
    R^{sp}(\Gamma) \;
    \sum_{\Pi\in\cP(\cL_\Gamma)}
   \; \left( \prod_{P\in\Pi} m^{1 c}_{|P|} \right)
   \;
  \left( \sum_{\Pi\to\overline{\Lambda}_1}^{\quad\prime}
  \prod_{l\in\overline{\Lambda}_1}
   1 \right) .
\ee
The last summation in (\ref{sg.wsptodlce}) is over all
maps $\cL_\Gamma\to\overline\Lambda_1$ of the lines
of $\Gamma$ to the lattice links of $\overline\Lambda_1$
subject to the constraint that all lines that belong
to the same multiple-line corresponding to some $P\in\Pi$
are mapped onto the same lattice link.

Finally we rewrite (\ref{sg.wsptodlce}) as a sum over
\mlgraphs.
To this end, we first observe that for every
$\Gamma\in\cG^{sp}_0(L)$, every partition
$\Pi\in\cP(\cL_\Gamma)$ of the lines of $\Gamma$ into 
multiple-lines generates a \mlgraph $\Delta=(\Gamma,\Pi)$ in the
obvious way.
Let us denote by
$\overline\cG_{0,0}(L)$ the subset of \mlgraphs of
$\cG_{0,0}(L)$ that stay connected after decomposition
of all multiple lines.
(These are the \mlgraphs which stay connected in the usual
graph theoretical sense, when the dashed lines are omitted.)
For every $\Delta\in\overline\cG_{0,0}(L)$ there is a unique
$\Gamma(\Delta)\in\cG_0^{sp}(L)$ and at least one
$\Pi\in\cP(\cL_{\Gamma(\Delta)})$ such that
$(\Gamma(\Delta),\Pi)=\Delta$.
Let $n_\Delta$ be the (uniquely determined) number of
partitions $\Pi\in\cP(\cL_{\Gamma(\Delta)})$ with
$(\Gamma(\Delta),\Pi)=\Delta$,
and $\Pi(\Delta)$ such an arbitrary partition.
Eq.~(\ref{sg.wsptodlce}) then becomes
\be \label{sg.wdlce}
  [[ \frac{1}{V} \; W_{sp}(U) ]]  =
   \sum_{L\geq 0} (2K)^L \sum_{\Delta\in\overline\cG_{0,0}(L)}
    n_\Delta R^{sp}(\Gamma(\Delta)) \;
   \; \left( \prod_{P\in\Pi(\Delta)} m^{1 c}_{|P|} \right) \;
  \left( \sum_{\Pi(\Delta)\to\overline{\Lambda}_1}^{\quad\prime}
   1 \right) .
\ee
The last bracket of (\ref{sg.wdlce})
is the lattice embedding factor of the
\mlgraph $\Delta$.
The second bracket from the right does not depend on the choice
of $\Pi(\Delta)$ and is the product of the multiple-line
coupling constants as defined in section 3.
Finally, $n_\Delta R^{sp}(\Gamma(\Delta))$ is precisely
the remaining part of the weight of $\Delta$
that was described in detail in section 3, endowed
with the correct inverse topological symmetry number
of the \mlgraph $\Delta$ (because of the factor $n_\Delta$).

In summary, we obtain the series expansion of the link-averaged
free energy density in terms of DLCE graphs,
\be \label{sg.wseries}
    [[ \frac{1}{V} \; W_{sp}(U) ]] \; = \;
    \sum_{L\geq 0} (2K)^L 
     \sum_{\Delta\in\overline{\cG}_{0,0}(L)} w(\Delta) .
\ee
The weight $w(\Delta)$ of a \mlgraph $\Delta$ is defined
and computed according to the rules given in section 3.

Eq.~(\ref{sg.wseries}) is the series representation of the
link-averaged free energy density of the spin system,
i.e.~the free energy density of the $n=0$ replica system,
in terms of DLCE graphs.
It looks much like the series representation of the
$1$-replica system, which is given by
\be \label{sg.w1series}
    \frac{1}{V} \; W_{1-repl} \; \equiv \;
    \frac{1}{V} \; \ln{[[ \exp{W_{sp}(U)} ]]} \; = \;
    \sum_{L\geq 0} (2K)^L 
     \sum_{\Delta\in\cG_{0,0}(L)} w(\Delta)
\ee
according to the discussion of section 2.
We recall that
$\cG_{0,0}(L)$ is the set of DLCE vacuum graphs with $L$ bare lines
that are connected in the generalized DLCE sense.
Comparing (\ref{sg.wseries}) and (\ref{sg.w1series}),
the transition from $n=1$ to $n=0$ replicas is achieved by
keeping only the subset
$\overline{\cG}_{0,0}(L)\subseteq\cG_{0,0}(L)$ of
\mlgraphs that are connected in the original (LCE) sense.

We expect that the series (\ref{sg.wseries}) are convergent
for a large class of interactions $S^1(U)$ and $v(x,y)$
if the coupling constant $K$ is sufficiently small.
For special interactions most of the \mlgraphs yield 
vanishing contributions so that we can further restrict
the sum to a subset of $\overline{\cG}_{0,0}(L)$.
An example is given by the mean field type of interaction
of the SK model
where all pairs of sites are coupled with the same strength.
That is, in the notation of this subsection,
\bea
v(x,y) & = & K (1-\delta_{x,y}),  \nonumber \\
\overline\Lambda_1 & \equiv & \Lambda_1 \;
\mbox{is the set of all pairs
of sites,} \\
S^1(U) & = & V \; \frac{1}{2} \; U^2 . \nonumber
\eea
In the thermodynamic limit $V\to\infty$ only those \mlgraphs
survive that consist exclusively of $2$-lines, and that do not contain
any loop of $2$-lines.
The resulting tree structure provides the possibility of summing
the series (\ref{sg.wseries}) and of analytically continuing
the result to the spin glass phase of large coupling $K$.
Work in this direction is in progress.
%
%
%
%
\section{Applications to the SU(2) Higgs model}

In this section we discuss applications of DLCEs to study the 
Higgs transition within a variational
estimate for the gauged SU(2) Higgs model.
The estimate serves as an effective
description of the electroweak standard model.

We consider a 4-dimensional hypercubic finite temperature lattice
$\Lambda_0$ of size $L_0\times V_3$, with $L_0=T^{-1}$ the inverse
temperature in lattice units and $V_3$ the spatial volume.
The lattice links are the set of nearest neighbour lattice sites,
given by
\be
   \overline\Lambda_1 \; = \; \{ (x;\mu) \; \vert \; x\in\Lambda_0,
   \mu=0,\dots ,3 \} .
\ee
The gauge field $U(x;\mu)$ is an SU(2) valued field living on the
lattice links $\overline\Lambda_1$.
It is convenient to parametrize such an SU(2) matrix by
\bea \label{su2.uparam}
   U & = & \phi_0(U) \; {\bf 1}_2 + i \; \vec{\sigma}\cdot \vec{\phi}(U)
  \nonumber \\ 
    & \equiv & U_0 + i \; \vec{\sigma}\cdot \vec{U}
\eea
with $U \simeq (U_0,\vec{U})\in S_3$.
The Higgs field $\Phi$ lives on the lattice sites $\Lambda_0$.
Its values are real multiples of SU(2),
\bea \label{su2.phiparam}
   \Phi(x) & = & \phi_0(\Phi(x)) \; {\bf 1}_2 
      + i \; \vec{\sigma}\cdot \vec{\phi}(\Phi(x))
  \nonumber \\ 
    & \equiv & \phi_0(x) + i \; \vec{\sigma}\cdot \vec{\phi}(x),
\eea
with $\phi \simeq(\phi_0(x),\vec\phi(x))\in{\bf R}^4$.
We say that $\Phi$ is $cU(2)$ valued.

As a special case of Eq.~(\ref{2.zgen}) we consider the partition function
\be\label{su2.zdlce}
  Z_{VE1} \; = \; \int\cD U \cD\Phi \;
  \exp{(-S_{VE1}(U,\Phi))},
\ee
in which
\be
  \cD U \; = \; \prod_{x\in\Lambda_0} \prod_{\mu=0}^3
  d\mu_H(U(x;\mu)),
\ee
with $d\mu_H(U)$ the normalized Haar measure on SU(2), and
\be \label{su2.phimeasure}
  \cD\Phi = \prod_{x\in\Lambda_0} \exp{(-S^\circ(\Phi(x)))}
   \; d\nu(\Phi(x)),
\ee
where
\bea
  d\nu(\Phi) & = & d^4\phi, \nonumber \\
  S^\circ(\Phi) & = &  \frac{1}{2} {\rm tr\,}{(\Phi^\dagger \Phi)}
  + \lambda \left( \frac{1}{2} {\rm tr\,}{(\Phi^\dagger \Phi)} 
    - 1 \right)^2
  \\
  & = & \phi^2 + \lambda \left( \phi^2 - 1 \right)^2 . \nonumber
\eea
Finally,
\bea\label{su2.svei}
  S_{VE1}(U,\Phi) & = & - \sum_{x\in\Lambda_0} \biggl\{
    4 \zeta_{link} \phi_0(U(x;0))
  + \sum_{\mu=1}^{3} 4 \zeta_{cube} \phi_0(U(x;\mu))
   \nonumber \\
  && \qquad + \sum_{\mu=1}^{3} (2\kappa)
     \frac{1}{2} {\rm tr\,}{(\Phi(x)^\dagger U(x;\mu)\Phi(x+\widehat\mu))}
   \\
  && \qquad  + \xi \phi_0(\Phi(x)) \biggr\} , \nonumber
\eea
depending on variational parameters $\zeta_{link}$, $\zeta_{cube}$ and $\xi$
$\in {\bf R}$, and on the hopping parameter $\kappa$.
The first two terms are non-gauge invariant substitutes
for the Wilson plaquette term of the gauge part of the SU(2) Higgs model,
cf. Eq.~(\ref{su2.1.wilson}) below. 
We distinguish timelike from spacelike directions.
Timelike links are denoted as $(x;0)$ and spacelike as $(x;\mu)$,
$\mu =1,2,3$. The third term is a hopping term coupling $\Phi$
and $U$ fields only in spatial directions.
The last term depends on the third variational parameter $\xi$
associated with the Higgs field $\Phi$, while the remaining ultralocal
action for $\Phi$ has been absorbed in the measure.

In the next subsection we first indicate how we arrive at the 
action (\ref{su2.svei}) within a variational estimate for the free energy 
of the  SU(2) Higgs model. We derive equations for the optimal choice
of variational parameters and an equation -based on stability arguments-
to determine the critical hopping parameter 
$\kappa_{crit}$ at which the Higgs phase
transition sets in. These equations will depend on expectation values
of n-point correlations that are most suitably evaluated with a DLCE in
three dimensions.

The generic graphical expansion exposed in sections 3-5 now has 
to be adapted to account for the internal symmetry. This is somewhat
involved so that we restrict our discussion to some remarks in
section 7.2, but postpone any details to a forthcoming paper 
\cite{hilde2}. In section 7.3 we present some results for the critical line in
comparision with other variational estimates and Monte Carlo simulations.

%
\subsection{Variational estimates for the free energy of the SU(2) Higgs
  model}

By means of DLCEs we want to obtain an analytic
estimate for the critical line in the SU(2) Higgs model that serves
as an effective description of the Higgs transition in the electroweak
standard model. Because of the rather different methodical ansatz in this
approach, for a common choice of parameters this estimate will 
provide an independent check of
numerical results based on Monte Carlo simulations. In addition, larger
scalar field couplings are available in our approach and larger values for
$L_0$, the number of time slices. (In our calculations $L_0=4$ or 
$L_0=\infty$, while it was chosen as 2 or 3 in  \cite{jansen}.) Larger
$L_0$s are required for a finite size scaling control.

To fix the notation we introduce the action of the SU(2) Higgs model as
\bea\label{su2.1.higgs}
Z \; &=& \; \int \cD U \cD \Phi \; \exp{(-S(U, \Phi))} , \nonumber \\
S(U, \Phi) \; &=& \; S_W(U) + S_{hop} , \\
S_{hop} & = & - \sum_{x\in\Lambda_0} \biggl\{
     \sum_{\mu=0}^{3} (2\kappa) 
     \frac{1}{2} {\rm tr\,}{(\Phi(x)^\dagger U(x;\mu)\Phi(x+\widehat\mu))}
    \biggr\} . \nonumber
\eea
The Higgs self-interaction
has been absorbed in the measure $\cD \Phi$ of 
Eq.~(\ref{su2.phimeasure}),
$S_{hop}$ is the hopping
parameter term in 4 dimensions.
$S_W$ is the SU(2) gauge invariant Wilson
action
\be\label{su2.1.wilson}
  S_W(U) \; = \; \sum_{x\in\Lambda_0} \sum_{\mu<\nu=0}^3
  \overline\beta \; \left( 1 - \frac{1}{2} {\rm Re\,} {\rm tr \,}
  U(x;\mu) U(x+\widehat\mu;\nu)
  U(x+\widehat\nu;\mu)^{-1} U(x;\nu)^{-1} \right) 
\ee
with gauge coupling $\overline{\beta}\equiv 4/g^2$ 
and $U \in SU(2)$ as in (\ref{su2.svei}).

\vskip7pt

A plaquette term involves a 4-link interaction. In contrast to the hopping
parameter term it does not allow for a direct treatment with DLCEs.
Therefore we replace $Z$, the full partition function of the SU(2) 
Higgs model, by a partition function $Z_{VE}$ that is related to $Z$
by an inequality of the form
\be\label{su2.1.var}
  \exp{(-V f)} \equiv Z \; \geq \; Z_{VE}
   \; \exp{( < -\left( S-S_{VE}(\zeta) \right) >_{VE} )}
   \; \equiv \; \exp{(-V \widetilde{f}(\zeta))} .
\ee
Eq.~(\ref{su2.1.var}) follows from the convexity of the exponential
function and holds independently of the specific choice of $Z$ and 
$Z_{VE}$, if the measure is positive definite and normalized.
$V$ is the 4-dimensional volume, $f$ denotes
the true physical free energy density defined via $Z$, $\widetilde{f}$
the trial free energy density defined via the second equality in
(\ref{su2.1.var}). $S$ refers to the original action in $Z$, in our case it
is the SU(2) Higgs action, and $S_{VE}$ to an ansatz for the action in
$Z_{VE}$ depending on a generic set of variational parameters $\zeta$.
These parameters should be optimized so that the difference of the
trial and the physical free energy density 
$\widetilde{f}(\zeta)-f\geq 0$ becomes minimal for 
$\zeta=\widetilde{\zeta}$ at which $\widetilde{f}$ takes its
minimum.

The most naive ansatz for $S_{VE}$ is a mean field ansatz in the spirit
of a molecular field approximation, leading to a factorization 
of $Z_{VE0}$ according to
\be\label{su2.1.zve0}
Z_{VE0}(\zeta,\xi) \; = \; Z_{link}^{4 L_0 V_3}(\zeta)
                    \cdot  Z_{site}^{L_0 V_3}(\xi).
\ee
The partition function depends on two variational parameters
$\zeta$ and $\xi$ and factorizes
in a product over single link ($Z_{link}$) and single site ($Z_{site}$)
partition functions, $V_3$ denotes the three-dimensional volume.
In this ansatz there is no space for implementing an asymmetry between
temporal and spatial directions. Thus the results will be temperature 
independent by construction.

One can think of a variety of improvements of the molecular
field ansatz, arguments against or in favour of the various versions
will be given in \cite{hilde2}. Here we only consider the ansatz 
which is most
plausible from a physical point of view and leads to the best agreement
with Monte Carlo results for comparable sets of parameters. The ansatz
treats the spacelike degrees of freedom of the hopping term in 3 
dimensions beyond the mean field level (that is with DLCE),
but the timelike
degrees of freedom within a mean field approach, so that the partition
function factorizes over the spacelike hyperplanes. The reason is that
the spatial hopping term is supposed to contain the nonperturbative
properties of the full model that drive the Higgs phase transition.
While a mean field approach for all variables will be 
too rough to produce
high quality results for $\kappa_{crit}$, it appears more reasonable for 
timelike variables, although a finite temperature effect on the Higgs
transition gets lost this way. (We should remark that from results in a
scalar $\Phi^4$ theory in four dimensions the finite temperature effect
on $\kappa_{crit}$ is expected to be anyway quite small
\cite{LW1,thomas2}.)

Now it is easily checked that the choice of $S_{VE1}$ in 
Eq.~(\ref{su2.svei}) leads to a factorization
of $Z_{VE1}$ according to
\be\label{su2.1.zvei}
   Z_{VE1}(\zeta_{link},\zeta_{cube},\xi) 
   \; = \; Z_{cube}(\zeta_{cube},\xi)^{L_0} 
   \cdot Z_{link}(\zeta_{link})^{L_0 V_3}.
\ee
The partition function $Z_{cube}$
of the spacelike degrees of freedom reads
\be\label{su2.1.zcube1}
   Z_{cube} \; = \; \int \prod_{x\in\Lambda_0^{(3)}} 
   \left(  d\nu(\Phi(x))
   {\rm e}^{-S^\circ(\Phi(x))}
    \prod_{\mu=1}^3 d\mu_H(U(x;\mu)) \right)
   \cdot \exp{(-S_{cube}(U,\Phi))} ,
\ee
with $\Lambda_{0}^{(3)}$ the 3-dimensional lattice and
\bea\label{su2.1.zcube2}
    S_{cube}(U,\Phi) & = & - \sum_{x\in\Lambda_0^{(3)}} \biggl\{
     \sum_{\mu=1}^{3} 4 \zeta_{cube} \phi_0(U(x;\mu))
      + \xi \phi_0(\Phi(x))
   \nonumber \\
  && \qquad + \sum_{\mu=1}^{3} (2\kappa) 
     \frac{1}{2} {\rm tr\,}{(\Phi(x)^\dagger U(x;\mu)\Phi(x+\widehat\mu))}
    \biggr\} , 
\eea
while $Z_{link}$ is an ultralocal one-link integral
\be\label{su2.1.z0}
   Z_{link} \; = \; \int d\mu_H(U) \; \exp{(-S_{link}(U))}
\ee
with
\be\label{su2.1.s0}
   S_{link}(U) \; = \; 4 \zeta_{link} \phi_0(U) .
\ee
Expectation values $<O>_{cube}$, $<O>_{link}$, 
$<O>_{VE1}$ of observables $O$ refer to $Z_{cube}$, 
$Z_{link}$ and $Z_{VE1}$, respectively.
Minimization of the trial free energy density
$\tilde{f}(\zeta_{link},\zeta_{cube},\xi)$
in terms of these expecation values leads to three equations. The first one
$\left(\partial\widetilde{f}/\partial\xi\right)=0$ is solved by $\xi=0$ in
the symmetric phase.
The remaining two equations are given by
\bea\label{su2.1.mfeq1}
  && \frac{\bar{\beta}}{4}
  \frac{\partial\widetilde{W}^{1,2}_{cube}}{\partial\zeta_{cube}}
  + \left\lbrack \bar{\beta} (< \phi_0(U) >_{cube})^3
       + \frac{\bar{\beta}}{2} (< \phi_0(U) >_{link})^2 < \phi_0(U) >_{cube}
       - \zeta_{cube} \right\rbrack
   \nonumber \\
   && \qquad\qquad \cdot \; \frac{\partial}{\partial\zeta_{cube}} 
     < \phi_0(U) >_{cube}
      \quad = \quad 0
\eea
and
\bea\label{su2.1.mfeq2}
  \left\lbrack \frac{3 \bar{\beta}}{2} (< \phi_0(U) >_{cube})^2
   < \phi_0(U) >_{link} - \zeta_{link} \right\rbrack
   \cdot \frac{\partial}{\partial\zeta_{link}} < \phi_0(U) >_{link}
  \quad = \quad 0.
\eea
Here we have used the shorthand notation
\bea\label{su2.1.plaq}
 < \phi_0(U) >_{cube} & \equiv & 
    < \phi_0(U(x;1)) >_{cube} , \nonumber \\
 \widetilde{W}_{cube}^{1 ,2} & \equiv &
   \frac{1}{2} < {\rm tr\,} U(x;1) U(x+\widehat{1};2)
   U(x+\widehat{2};1)^{-1} U(x;2)^{-1} >_{cube}
 \\
  & - &  (<\phi_0(U(x;1))>_{cube})^4, \nonumber
\eea
with $x\in\Lambda_0^{(3)}$.
In (\ref{su2.1.plaq}) we have used the lattice symmetries.
Eq.s~(\ref{su2.1.mfeq1}), (\ref{su2.1.mfeq2}) should be solved for 
$\zeta_{link}$, $\zeta_{cube}$, and $\xi$ as series in $\kappa$.
The stability condition for the symmetric minimum at
$\xi=0$ is given as
\bea\label{su2.1.stab}
  && \quad 3 \overline\beta \;
     \frac{\partial^2\widetilde{W}^{1,2}_{cube}}{\partial\xi^2}
  \bigg\vert_{\xi=0} \nonumber \\
  && + 12 \left\lbrack \bar{\beta} (< \phi_0(U) >_{cube})^3
 + \frac{\bar{\beta}}{2}
 < \phi_0(U) >_{cube} (< \phi_0(U) >_{link})^2 - \zeta_{cube} \right\rbrack
  \nonumber \\
  && \qquad \cdot \; \frac{\partial^2}{\partial \xi^2} 
   < \phi_0(U) >_{cube} 
  \bigg\vert_{\xi=0}  \\
  && + \left\lbrack (2\kappa) \, 2 \, < \phi_0(U) >_{link}
    \frac{\partial}{\partial\xi} <\phi_0(\Phi) >_{cube} \bigg\vert_{\xi=0}
   - 1 \right\rbrack
     \cdot \; \frac{\partial}{\partial\xi} < \phi_0(\Phi) >_{cube}
  \bigg\vert_{\xi=0} \nonumber \\
  && \nonumber \\
  && \; < \; 0, \nonumber
\eea
with
\be
  < \phi_0(\Phi) >_{cube} \; \equiv \; 
  < \phi_0(\Phi(x)) >_{cube},
\ee
$x\in\Lambda_0^{(3)}$.
For an equality sign, (\ref{su2.1.stab}) determines $\kappa_{crit}$ order by
order in the expansion. 

Note the two type of expectation values $<\cdot>_{link}$, $<\cdot>_{cube}$
entering Eq.s~(\ref{su2.1.mfeq1}), (\ref{su2.1.mfeq2}), (\ref{su2.1.stab}).
While the single link expectation values $<\cdot>_{link}$ can be evaluated
exactly, the $<\cdot>_{cube}$s must be approximated. The derivatives of
$\widetilde{W}^{1,2}_{cube}$ w.r.t. $\zeta_{cube}$ or twice w.r.t.
$\xi$ induce up to 5-point functions in $U$ and 6-point functions in
four $U$s and two $\Phi$s. Because of the bad signal/noise ratio Monte
Carlo calculations of such connected correlations would
not be feasible.
Therefore this is the place, where we use a DLCE to evaluate expectation
values of the type $<U\cdots U \Phi\cdots \Phi>_{cube}$ (all internal and 
configuration space indices suppressed) order by order in $\kappa$.
Clearly, also in this scheme the price is high. For example, the number of
connected DLCE graphs contributing to order $\kappa^4$ to a 4-point function
of 2 $U$s and 2 $\Phi$s is about 100. Not all correlations in
(\ref{su2.1.mfeq1}),(\ref{su2.1.mfeq2}),(\ref{su2.1.stab}) appear in the
form of susceptibilities. In a product of 4 $U$s, for instance, the 
configuration space indices are fixed to the boundary of a plaquette. Such
features must be noticed for calculating the graphical weights.

These remarks may indicate the complexity of the actual evaluation
of Eq.s~(\ref{su2.1.mfeq1}), (\ref{su2.1.mfeq2}), (\ref{su2.1.stab}).
Here we omit any further details on the list of contributing graphs and
postpone them to \cite{hilde2}. Before we quote some results, we
comment on further subtleties, when DLCEs are 
adapted to internal symmetries of the hopping term.

%
\subsection{DLCEs for internal symmetries}

In this section we focus on the (gauge) group $SU(2)$,
since we are interested
in applications to the electroweak phase transition, but 
it should be obvious,
how one could proceed along the same lines for a group 
$SU(N)$ with $N>2$.
In the development of the \mlgraph theory in sections 3-5
we had assumed that $\phi \in {\bf R}$, $U \in {\bf R}$ to simplify
the notation. In Eqs.~(\ref{su2.uparam}) and (\ref{su2.phiparam}) 
we had $\Phi \in cU(2)$, $U \in SU(2)$.
Let us write the hopping part of the action in the symmetrized form
 
\be\label{su2.hop1}
   S_{hop} \; = \; - \frac{1}{2} \sum_{x,y\in\Lambda_0}
   v(x,y) \; \frac{1}{2} \; {\rm tr}\, \Phi(x)^\dagger U(x,y) \Phi(y) ,
\ee
with hopping parameter
\be
   v(x,y)  =  2\kappa \; \sum_{\mu=0}^3
    \left( \delta_{y,x+\widehat\mu} + \delta_{y,x-\widehat\mu} \right)
\ee
and
\be
    \begin{array}{c@{\quad =\quad}c}
    U(x,x+\widehat\mu) & U(x;\mu) \\
    U(x+\widehat\mu,x) & U(x;\mu)^{-1} \\
    U(x,y) & 0 \quad {\rm otherwise}
    \end{array} 
   \; , \;\mu=0,\dots ,3, 
\ee
Using the parametrizations
(\ref{su2.uparam}) and (\ref{su2.phiparam})
we have
\bea\label{su2.hop2}
    && \frac{1}{2} {\rm tr}\, \Phi(x)^\dagger U(x,y) \Phi(y) 
   \; = \; \phi(x) \cdot \phi(y) \; U_0(x,y)
  \nonumber \\
   && \qquad + \phi_0(y) \; \vec{\phi}(x) \cdot \vec{U}(x,y)
             - \phi_0(x) \; \vec{\phi}(y) \cdot \vec{U}(x,y)
  \\
  && \qquad + \vec{\phi}(x) \times \vec{\phi}(y) \cdot \vec{U}(x,y).
  \nonumber
\eea

Note that in the parametrization in terms of $(\phi_0, \vec{\phi})$s
and $(U_0,\vec{U})$s the $S^\circ$ part of the action is O(4) invariant,
but $S_{VE1}$ is only O(3) invariant for $\zeta \neq 0$ or $\xi \neq 0$.
Thus we have to distinguish between singlet $(0)$ and triplet $(s=1,2,3)$
indices in internal space.

Recall the generating equations (\ref{2.genequation}) of the 
graphical expansion of 
a DLCE in which $<\Phi(x) U(x,y) \Phi(y)>$ had been expressed in terms
of connected $n$-point functions, $n=1,2,3$. Using the $O(3)$ symmetry
it follows from Eq.~(\ref{su2.hop2})
that each derivative $\partial /\partial I(x,y)$ now has to be 
replaced by the differential operator
\be\label{su2.lab}
   L_{ab}(x,y) \; = \; \delta_{ab} \frac{\partial}{\partial I_0(x,y)}
   + \sum_{\gamma=1}^3 \left( \delta_{a,\gamma} \delta_{b,0}
   - \delta_{b,\gamma} \delta_{a,0} + \epsilon_{ab\gamma}
     \right) \frac{\partial}{\partial I_\gamma(x,y)}.
\ee
Latin indices run from $0$ to $3$, Greek indices from $1$ to $3$. 
Furthermore we have introduced the 
totally antisymmetric $\epsilon$-symbol which is zero if any
of the three indices is zero and $\epsilon_{123}=1$.
This implies for instance for the
one nonvanishing term at $v=0$ in Eq.~(\ref{2.genequation}) a substitution
according to
\bea\label{su2.ww}
W_{H(x)}W_{H(y)}W_{I(x,y)} \; \longrightarrow \; W_{H^a(x)}W_{H^b(y)}
         L_{ab}(x,y)W.
\eea
According to the graphical rules an $n$-line was generated by
$n$ iterated derivatives w.r.t.
$I_j, j \in 0,1,2,3$, evaluated at $v=0$.
These derivatives now have to be replaced by a
multiple application of $L$ yielding an n-line

\be \label{su2.Labgraph}
{
\setlength{\unitlength}{0.8cm}
%
%
\begin{picture}(5.0,1.0)

%

\put(0.0,-0.6){
\setlength{\unitlength}{0.8cm}
\begin{picture}(4.0,0.0)

\qbezier(1.0,1.5)(2.0,2.1)(3.0,1.5)
\qbezier(1.0,1.0)(2.0,1.6)(3.0,1.0)
\qbezier(1.0,0.0)(2.0,0.6)(3.0,0.0)

\put(-0.2,0.7){\makebox(0.2,0){$x$}}
\put(4.2,0.7){\makebox(0.2,0){$y$}}

\put(0.6,1.5){\makebox(0.2,0){$a_1$}}
\put(3.3,1.5){\makebox(0.2,0){$b_1$}}

\put(0.6,1.0){\makebox(0.2,0){$a_2$}}
\put(3.3,1.0){\makebox(0.2,0){$b_2$}}

\put(0.6,0.0){\makebox(0.2,0){$a_n$}}
\put(3.3,0.0){\makebox(0.2,0){$b_n$}}

\put(1.6,1.1){\makebox(0.2,0){$\cdot$}}
\put(1.6,0.8){\makebox(0.2,0){$\cdot$}}
\put(1.6,0.5){\makebox(0.2,0){$\cdot$}}

\put(2.0,1.85){\line(0,1){0.2}}
\put(2.0,1.6){\line(0,1){0.2}}
\put(2.0,1.35){\line(0,1){0.2}}
\put(2.0,1.1){\line(0,1){0.2}}
\put(2.0,0.85){\line(0,1){0.2}}
\put(2.0,0.6){\line(0,1){0.2}}
\put(2.0,0.35){\line(0,1){0.2}}
\put(2.0,0.1){\line(0,1){0.2}}

\end{picture}
}
%

\end{picture}
}
   \; = \;
   \sum_{\mu=0}^3 \left( \delta_{y,x+\widehat\mu} +
   \delta_{y,x-\widehat\mu} \right) \;
   (2\kappa)^n
   \left. \left( \prod_{i=1}^n L_{a_ib_i} \right) 
    W^1(I) \right\vert_{I=0}
\ee
where the $1$-link generating function $W^1(I)$ is defined by
\be
  \exp{ W^1(I) } \; = \;
  \frac{ \int d\mu_H(U) \; \exp{( 4\zeta_{cube}\phi_0(U) 
    + \sum_{a=0}^3 I_a \phi_a(U) )}
  }{ \int d\mu_H(U) \; \exp{( 4\zeta_{cube}\phi_0(U) )}
  } .
\ee

For $a_i=b_i=1$ for all $i \in 1,...,n$, $n$ applications of $L_{ab}$ to
$W^1$ reduce to $\delta_{ab} \partial^n W^1/\partial I^n \mid_{I=0}$, 
but otherwise
we get a complicated mixing of singlet and triplet terms.

\begin{figure}[h]

\begin{center}
\setlength{\unitlength}{0.8cm}

%
%
\begin{picture}(15.0,4.0)

%
%
%

\epsfig{bbllx=-333,bblly=263,
        bburx=947,bbury=673,
        file=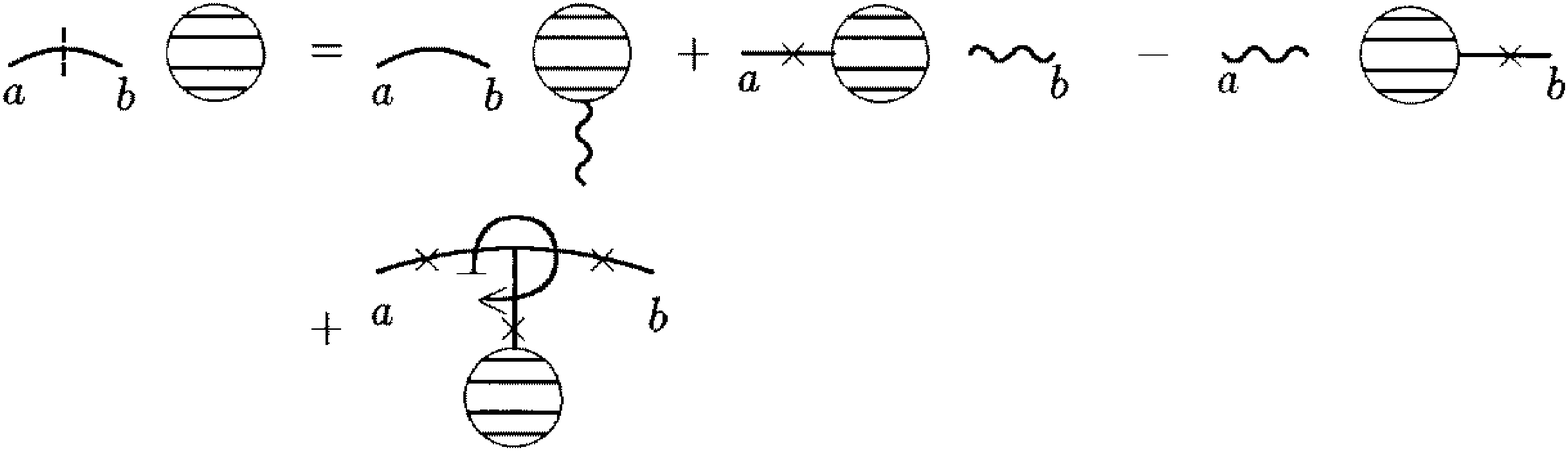,
        scale=0.27}

%
%

\end{picture}
%
%
\end{center}

\caption{\label{Lab} Graphical representation of $L_{ab}W^1$.
It provides the generating equation for the graphical computation
of the internal symmetry factor of a multiple-line.
It replaces Fig.~1(b) for $I=H=0$, $v=0$ and $n=1$.
For further explanations see the text.}
\end{figure}

To evaluate products of $L$s systematically, 
it is again quite convenient to manipulate graphical
rather than analytic expressions. We represent one application of 
$L_{ab}$ on $W^1$ as shown in
Fig.~(\ref{Lab}). Multiple applications of $L_{ab}$ require an iteration of
Fig.~(\ref{Lab}).


Twiddle lines are bare $\phi$-lines with a singlet index $a=0$, crossed
lines are bare $\phi$-lines with a triplet index $a =1,2,3$.
The arrow in the last term stands for the
$\epsilon_{ab\gamma}$ symbol
in $L_{ab}$, its orientation reflects the order of indices $ab\gamma$.

This way one achieves a further graphical "decomposition" of an 
n-line into terms merely consisting of bare lines, from which the 
internal symmetry
factors can be easily read off. For
example a 2-line is further decomposed into 11 terms, a 3-line into
29 and a 4-line into 142. 
This may indicate that the number of decomposed contributions
rapidly increases with the number $n$ of bare lines collected to an n-line.
In our actual calculations up to order $\kappa^4$ the maximal number 
of internal lines was 4 and
the highest multiple line was a 6-line with 4 internal and 2 
external U-lines. Some of the results will be reported in the next section.

%
\subsection{Some results}

In Table 1 we compare results for $\kappa_{crit}$ obtained for 2 values of
$\bar{\beta}$ and $\lambda$ and various lattice volumes 
within three methods.
The first line refers to a naive molecular field approximation with a
partition function $Z_{VE0}$ factorizing according to
Eq.~(\ref{su2.1.zve0}).
Because of the complete
factorization in space and time directions the ansatz keeps only a trivial
volume dependence and is temperature independent by construction.

The second line refers to Monte Carlo simulations of the SU(2) Higgs
model in four dimensions \cite{jansen} for two lattice extensions
$L_0=2$ and $3$. 

The third line shows results of the DLCEs with the variational ansatz
(VE1) presented in section 7.2.
Again this ansatz is independent of $L_0$ by construction, so that we
get an identical result for $L_0= \infty$. The values for $\kappa_{crit}$
have been obtained by a linear regression in $1/R^2$, $R$ denoting the
order in $\kappa$ of the DLCE. For the regression we used the
coefficients of $\kappa^1$, $\kappa^2$, $\kappa^3$ and $\kappa^4$.
The value of $\kappa_{crit}$ at order $\kappa^4$ was given by $0.1284$
for $\lambda=5\cdot 10^{-4}$ and $\overline\beta = 8.0$.
Although the fourth order seems to be low for a series expansion, the
agreement with the Monte Carlo results is not too surprising, if we
remind the complexity of the expansion, which is manifest in the number
of graphs contributing at fourth order to the various expectation values.

Furthermore, for otherwise fixed parameters, we see a decrease
of $\kappa_{crit}$
between the Monte Carlo and the DLCE results  
for increasing $L_0$. This is in qualitative agreement
with the expectation.


\begin{table}[htb]
\vspace{0.5cm}

\begin{center}
\begin{tabular}{|r|c|c|l|l|}
\hline \hline
  Method & $\bar{\beta}$ & $\lambda$ & lattice & $\kappa_{crit}$
  \\ [0.5ex] \hline
  Mean Field & $8.0$ & $5.0\cdot 10^{-4}$ & $\infty^4$ & $0.12973$
    \\
  Monte Carlo & $8.0$ & $5.0\cdot 10^{-4}$ & $2\cdot 32\cdot 32 \cdot 256$ & $0.12887(1)$
    \\
  VE1 \& DLCE & $8.0$ & $5.0\cdot 10^{-4}$ & $4\cdot \infty^3$ & $0.1282(1)$
    \\ \hline \hline
 Mean Field & $8.15$ & $5.1\cdot 10^{-4}$ & $\infty^4$ & $0.12964$ 
    \\
 Monte Carlo & $8.15$ & $5.1\cdot 10^{-4}$ & $3\cdot 48\cdot 48\cdot 384$ & $0.12852(2)$
    \\
 VE1 \& DLCE  & $8.15$ & $5.1\cdot 10^{-4}$ & $4\cdot \infty^3$ & $0.1281(1)$
    \\ \hline
\end{tabular}

\end{center}

\caption{\label{kc.1} Results for $\kappa_{crit}$ for 2 values of
$\overline\beta$ and $\lambda$, various lattice extensions,
and 3 methods. The Monte Carlo results are taken from [15],
the mean field and DLCE results, extrapolated to infinite order,
from this paper.
(The error in the DLCE results refers to the extrapolation of $\kappa_{crit}$.)
}
\end{table}

In Table 2 we compare results for $\kappa_{crit}$, to order $\kappa^4$ (first
column) and extrapolated to infinite order (with the same regression as
explained above) (second column) for DLCEs with different variational
estimates. The couplings are fixed, and $L_0=4$. VE1 refers to the first
variational ansatz, also exposed in
Table 1. The ans\"atze for the variational estimates VE2 and VE3 will
be explained in detail in \cite{hilde2}. Here we only characterize them by
their different factorization properties of the partition function.


\begin{table}[htb]
\vspace{0.5cm}

\begin{center}
\begin{tabular}{|l|l|l|}
\hline \hline
  Ansatz & $O(\kappa^4)$ & extrapolated
  \\ [0.5ex] \hline
  VE1 & $0.1284$ & $0.1282(1)$
    \\
  VE2 & $0.12958$ & $0.12910(6)$
    \\
  VE3 & $0.12948$ & $0.12934(7)$
    \\ \hline
\end{tabular}

\end{center}

\caption{\label{kc.2} Results for $\kappa_{crit}$, obtained in a
DLCE to order $\kappa^4$ (second column) and extrapolated to infinite order
(third column) at
$\beta=8.0$, $\lambda=5\cdot 10^{-4}$, $L_0=4$.
The ans\"atze VE1, VE2, VE3 refer to 3 variational estimates for the
free energy of an SU(2) Higgs model. 
Their gap equations are solved with DLCE.}
\end{table}

The partition function $Z_{VE2}$ factorizes acording to
\be\label{su2.zveii}
    Z_{VE2}(\zeta_{(0,string)},\zeta_{(s,link)},\xi )\; = \;
    Z_{(0,string)}(\zeta_{(0,string)},\xi)^{V_3}
    \cdot
    Z_{(s,link)}(\zeta_{(s,link)})^{3 L_0 V_3} ,
\ee
i.e., a product of a partition function $Z_{(0,string)}$ that couples
fields with hopping parameter terms 
along a string in time direction, and $Z_{(s,link)}$, a partition
function for a single spatial link. While expectation values w.r.t.
$Z_{(s,link)}$ can be evaluated exactly (within the numerical accuracy),
expectation values w.r.t. $Z_{(0,string)}$ are preferably
evaluated with DLCE.
This way they do depend on $L_0$, in principle they are sensitive 
to detect a temperature effect on $\kappa_{crit}$.
So far the couplings along a string are only kept in time direction, while
the spacelike directions are incorporated on a mean field
level in $Z_{(s,link)}$.

The third variational ansatz VE3 therefore is 
chosen as a symmetrized version of
VE2 in that the string now can extend in any of the 4 directions.
Again we treat space-and timelike degrees of freedom differently,
because we are interested in systems at finite temperature with
temperature dependent coupling $\kappa_{crit}$. The number of variational
parameters is larger, each of it depending on two indices telling us whether
the parameter belongs to a single link action or to a string, and whether
the string extends in temporal or spacelike directions.

For both versions VE2 and VE3 the actual $L_0$ dependence ($L_0=4,
\infty$) lies outside the systematic error, if we expand the series 
to order $\kappa^4$. For $L_0=4$ the fourth order is the minimal order 
in $\kappa$ at which a finite temperature effect is
visible in principle.
(For the graphs it should be possible to wind at least
once around the torus in $0$-direction.)
At O$(\kappa^4)$ the part of graphs for which the lattice embedding is
sensitive to $L_0$ is about $1\%$ of the total number of graphs. Such
a small "signal$/$noise" ratio produces an effect  
in the 7th or 8th digit, outside the error of the $O(10^{-4})$ because of
the truncation of the DLCE series. By increasing the order of
the expansion, finite-temperature effects will become more pronouced.
For higher orders an algorithmic implementation of the proliferating graphs
becomes unavoidable.
%
%
%
%
\section{Summary and Conclusions}

In this paper we have introduced a new expansion scheme for 3-point
interactions or, more precisely, for point-link-point interactions.
This scheme generalizes linked cluster expansions
for 2-point interactions 
by including hopping parameter terms endowed with their own dynamics.
In chapters 3-5 we have developed a \mlgraph theory with an
additional new type of multiple-line connectivity.
We have introduced appropriate equivalence classes of graphs
and discussed the issue of renormalization.
The main building blocks for an algorithmic generation of graphs
have been constructed.
Because of
the fast proliferation of graphs already at low orders in the expansion,
a computer aided implementation becomes unavoidable,
if one is interested in higher orders of the expansion than 
we have computed so far.

In chapter 6 we have indicated promising applications to spin glass 
systems. In particular we have identified the DLCE graphs contributing
to the link-average $[[\ln{Z_\beta(U)}]]$ of the spin free energy
$\ln{Z_\beta(U)}$. In the past such quantities often were only accessible
by means of the replica trick. 

In section 7 we have
outlined encouraging results for the transition line of the electroweak phase
transition within the SU(2)-Higgs model.
The results have been obtained by DLCEs applied
to gap equations that follow from
variational estimates of the free energy. 
They are in good agreement with corresponding high precision 
Monte Carlo results.
As we have seen, the main complications of DLCEs for SU(N) Higgs
systems come from the internal symmetry structure of the hopping term.
In a forthcoming paper we will demonstrate how one has to refine 
the graphical
representation to calculate internal symmetry factors merely within a
graphical expansion.

In conclusion, DLCEs are involved, but practicable, 
at least with computer aided
generation of graphs. They provide an analytical tool to study systems
in situations in which it has been impossible so far.

\section*{Acknowledgment}

We would like to thank Reimar K\"uhn (Heidelberg) for pointing out to
us ref.s \cite{sherrington}-\cite{penney}.
%
%
%
%


\end{document}